\documentclass[fleqn,usenatbib]{mnras}

\usepackage{sansmath}
\usepackage{makecell}
\usepackage{color}
\usepackage{bm,cancel}
\usepackage{booktabs}
\usepackage[normalem]{ulem}
\usepackage{widetext}

\newcommand{\expnumber}[2]{{#1}\mathrm{e}{#2}}
\newcommand{\rst}{\textit{Roman Space Telescope}}

\newcommand{\Ndist}[3]{ \mathcal{N}({#1},\,\expnumber{{#2}}{{#3}}) }
\newcommand{\CL}{\textsc{CosmoLike}\ }
\newcommand{\dd}{\mathrm{d}}
\newcommand{\mrm}[1]{\mathrm{#1}}
\newcommand{\mat}[1]{\textsf{\textbf{#1}}}
\newcommand{\Ha}{$\mathrm{H}\,\alpha$}
\newcommand{\OIII}{[O\,\textsc{III}]}
\newcommand{\bigo}[1]{\mathcal{O}(#1)}

\definecolor{orange}{rgb}{1.0, 0.49, 0}

\usepackage{scalerel}
\usepackage{tikz,xcolor}
\usetikzlibrary{svg.path}
\definecolor{lime}{HTML}{A6CE39}
\DeclareRobustCommand{\orcidicon}{%
	\begin{tikzpicture}
	\draw[lime, fill=lime] (0,0) 
	circle [radius=0.16] 
	node[white] {{\fontfamily{qag}\selectfont \tiny ID}};
	\draw[white, fill=white] (-0.0625,0.095) 
	circle [radius=0.007];
	\end{tikzpicture}
	\hspace{-2mm}
}
\foreach \x in {A, ..., Z}{%
	\expandafter\xdef\csname orcid\x\endcsname{\noexpand\href{https://orcid.org/\csname orcidauthor\x\endcsname}{\noexpand\orcidicon}}
}

\usepackage[T1]{fontenc}

\DeclareRobustCommand{\VAN}[3]{#2}
\let\VANthebibliography\thebibliography
\def\thebibliography{\DeclareRobustCommand{\VAN}[3]{##3}\VANthebibliography}

\usepackage{graphicx}	% Including figure files
\usepackage{amsmath}	% Advanced maths commands
\usepackage{amssymb}	% Extra maths symbols

\title[Kinematic Lensing with the Roman Space Telescope ]{Kinematic Lensing with the Roman Space Telescope}

\author[Xu et al.]{Jiachuan Xu\orcidA{},$^{1}$\thanks{E-mail: jiachuanxu@email.arizona.edu}
Tim Eifler,$^{1}$ 
Eric Huff,$^{2}$ 
Pranjal R. S.\orcidB{},$^{1}$ 
Hung-Jin Huang,$^{1}$
\newauthor
Spencer Everett,$^{2}$ 
Elisabeth Krause$^{1,3}$
\\
$^{1}$ Department of Astronomy/Steward Observatory, University of Arizona, 933 North Cherry Avenue, Tucson, AZ 85721-0065, USA \\
$^{2}$ Jet Propulsion Laboratory, California Institute of Technology, Pasadena, CA 91109, USA\\
$^{3}$ Department of Physics, University of Arizona, 1118 E Fourth Str, AZ 85721, USA}

\date{Accepted XXX. Received YYY; in original form ZZZ}

\pubyear{2021}

\begin{document}
\label{firstpage}
\pagerange{\pageref{firstpage}--\pageref{lastpage}}
\maketitle

%This is a simple template for authors to write new MNRAS papers.
%The abstract should briefly describe the aims, methods, and main results of the paper.
%It should be a single paragraph not more than 250 words (200 words for Letters).
%No references should appear in the abstract.
\begin{abstract}
Kinematic lensing (KL) is a new cosmological measurement technique that combines traditional weak lensing (WL) shape measurements of disc galaxies with their kinematic information. Using the Tully-Fisher relation KL  breaks the degeneracy between intrinsic and observed ellipticity and significantly reduces the impact of multiple systematics that are present in traditional WL. We explore the performance of KL given the instrument capabilities of the \rst, assuming overlap of the High Latitude Imaging Survey (HLIS), the High Latitude Spectroscopy Survey (HLSS) over 2,000 deg$^2$. Our KL suitable galaxy sample has a number density of  $n_{\mrm{gal}}=4~\mrm{arcmin}^{-1}$ with an estimated shape noise level of $\sigma_{\epsilon}=0.035$. We quantify the cosmological constraining power on $\Omega_{\mrm{m}}$-$S_8$, $w_p$-$w_a$ by running simulated likelihood analyses that account for redshift and shear calibration uncertainties, intrinsic alignment and baryonic feedback. Compared to a traditional WL survey we find that KL significantly improves the constraining power on $\Omega_{\mrm{m}}$-$S_8$ (FoM$_{\mrm{KL}}$=1.70FoM$_{\mrm{WL}}$) and $w_p$-$w_a$ (FoM$_{\mrm{KL}}$=3.65FoM$_{\mrm{WL}}$). We also explore a ``narrow tomography KL survey'' using 30 instead of the default 10 tomographic bins, however we find no meaningful enhancement to the FoM even when assuming a significant time-dependence in our fiducial dark energy input scenarios. 
\end{abstract}


%%%%%%%%%%%%%%%%%%%%%%%%%%%%%%%%%%%%%%%%%%%%%%%%%%

%%%%%%%%%%%%%%%%% BODY OF PAPER %%%%%%%%%%%%%%%%%%

\section{Introduction}
\label{sec:intro}

Weak lensing (WL) is one of the most promising cosmological probes since it directly traces the integrated gravitational potential along the line-of-sight (LoS) without any assumption on the relation between luminous matter and dark matter~\citep[see][for reviews]{Mandelbaum18,Kilbinger15,WMM13,BS01_review}. 
It plays an essential role in ongoing surveys such as 
the Dark Energy Survey (DES\footnote{\href{https://www.darkenergysurvey.org}
{\nolinkurl{https://www.darkenergysurvey.org}}}), 
the Kilo-Degree Survey (KiDS\footnote{\href{http://www.astro-wise.org/projects/KIDS/}
{\nolinkurl{http://www.astro-wise.org/projects/KIDS/}}}), the
Hyper Suprime Cam Subaru Strategic Program (HSC\footnote{\href{http://www.naoj.org/Projects/HSC/HSCProject.html}
{\nolinkurl{http://www.naoj.org/Projects/HSC/HSCProject.html}}}), 
and future surveys such as the
Rubin Observatory's Legacy Survey of Space and Time (LSST\footnote{\href{https://www.lsst.org}
{\nolinkurl{https://www.lsst.org}}}), the
\rst\footnote{\href{https://roman.gsfc.nasa.gov}
{\nolinkurl{https://roman.gsfc.nasa.gov}}} and 
\textit{Euclid}\footnote{\href{https://sci.esa.int/web/euclid}
{\nolinkurl{https://sci.esa.int/web/euclid}}}.
The currently ongoing, so-called Stage-III weak lensing surveys have achieved constraints on $S_8\equiv\sigma_8(\Omega_m/0.3)^{0.5}$ at the 3--5 per cent precision level. 
Constraints on $w_0$ range around a $40$ per cent precision level  using cosmic shear only~\citep{DES_Y3_WL2,DES_Y3_WL1,SubaruHSC_shear19,KiDS_shear20,KiDS+VIKING450xDES-Y1_shear}. 
This precision has been improved by combining cosmic shear with other probes such as galaxy galaxy lensing, galaxy clustering, CMB lensing, Supernovae, baryon acoustic oscillation (BAO), redshift space distortion (RSD) etc~\citep{DES_Y1_3x2pt,DESY1xPlanckxSPT_6x2pt,KiDS1000_3x2pt}. 
The near-future Stage-IV surveys are going to push this frontier to sub per cent level on $S_8$ and per cent level on the equation-of-state (EoS) of dark energy~\citep{WFIRST_AFTA,LSST_SRD18,Euclid_DSR,SKA_RedPaper}. 

However, weak lensing analyses from next generation galaxy surveys face several challenges. Firstly, standard weak lensing measurements require a large galaxy sample that includes predominantly faint galaxies in order to overcome limitations from intrinsic ellipticity noise.  
Secondly, photometric redshift (photo-$z$) estimation and shear calibration are inherently hard for faint galaxies and pose severe systematic uncertainties~\citep[e.g.,][]{Mandelbaum18} that must be modelled and controlled in future surveys. 
In order to overcome the first limitation the following novel methods have been proposed to directly infer the intrinsic galaxy shape:

    i) \textit{Radio polarization}: Observations have revealed azimuthal magnetic fields across the disc in the Milky Way and nearby spiral galaxies~\citep[e.g.~][]{HMQ_99,Beck07}. 
    When the galaxy is not perfectly aligned face-on towards the observer, Faraday rotation caused by the LoS component of the galactic magnetic field and free electrons will change the position angle (PA) of polarized synchrotron radiation.
    This results in inclination-dependent integrated polarization when the galaxy is not spatially resolved~\citep[][]{SKBT_09,SR12}. 
    Since gravitational lensing preserves the position angle, \cite{BB_11a} proposed to utilize polarization information to calibrate the intrinsic inclination of spiral galaxies. 
    However, typically only few per cent of the radiation is polarized, thus $\mu$Jy sensitivity is required to ensure a large enough sample with meaningful SNR.
    The relation between inclination and polarization as a function of environments and redshifts are further uncertainties in this method. Potentially, SKA will enable these type of measurements ~\citep{BB_11b,WBB_15,SKA_WL_3}.
    
    ii) \textit{Galaxy rotation velocity field}: This method was first proposed by~\cite{Blain02} and~\cite{Morales06}. 
    The idea is described as follows: weak lensing only linearly distorts the 2D position of photons, but keeps the frequency unchanged. 
    By measuring the projected 2D rotation velocity field of late-type galaxies, we can partially constrain the shapes before lensing assuming a disc kinematics model (see Sect.~\ref{sec:KL} for detailed description). 
    The rotation velocity could be measured both in optical/NIR bands traced by nebular emission lines, and in radio/submm wavelength traced by the CO rotational ladder or H\,\textsc{i} 21 cm emission. To fully break the degeneracy, one still needs information on either the direction of the shear field or the circular velocity of the disc. 
    \cite{DSM_15b,DSM_15a} proposed to estimate the direction of the shear field from the orientation of lens galaxies or clusters, and \cite{GTF20} obtained the first measurement from low redshift galaxy-galaxy lensing systems, with an averaged measurement of tangential shear $\langle \gamma_\mathrm{t} \rangle=0.020\pm0.008$ over 18 low-$z$ galaxies. While \cite{GTF20} primarily used the velocity map for their measurements, \cite{DiGiorgio21} further improved this approach by modelling the major and minor axes in photometry images of source galaxies. 
    \cite{Eric13,SKH+22} developed kinematic lensing in the context of cosmic shear rather than galaxy-galaxy lensing. Specifically, they built estimators for both shear components taking advantage of scaling relation of galaxies, e.g. the Tully-Fisher relation\citep[TFR,][]{TF1997}, to infer the circular rotation velocity of spiral galaxies. This will not only suppress the shape noise, but also decrease other systematic uncertainties like redshift error, shear calibration bias, and intrinsic alignment (see Sect.~\ref{sec:sys}). These improvements not only increase the cosmological constraining power of a tomographic WL survey, but also enable meaningful BAO detection from 2D tomography shear power spectra~\citep{DSH+19}. We note that synthesising methods of radio/optical/NIR rotation curves and radio polarization can produce more robust and precise measurements ~\citep{SKA_WL_3,BHH_18}.

Among the next generation telescopes the \rst~\citep{WFIRST_AFTA,Roman_general} has the most promising instrument capabilities, specifically high resolution grism and wide-field imaging, to conduct a self-contained KL survey. 
The ESA/NASA \textit{Euclid} mission could in principle also perform a KL measurement, the only concern is the slightly lower grism resolution and related uncertainty in measuring kinematics. 

The \rst\ has several primary science goals ranging from the search of habitable exoplanets to the origin and evolution of our universe. 
It will perform multiple surveys including a microlensing survey, exoplanets direct imaging, a type-Ia supernovae (SNIa) survey and, the main focus of this paper, a wide-field imaging and spectroscopic survey, the so-called High Latitude Survey (HLS). 
The HLS is designed to combine multiple cosmic probes, e.g. weak lensing, photometric galaxy clustering, galaxy clusters, BAO and RSD to optimize systematics control \citep[c.f.][]{WFIRST_MultiProbe}. 

In the current 1.6 year \textit{HLS reference survey design} the HLS is composed of the High Latitude Imaging Survey (HLIS) and High Latitude Spectroscopy Survey (HLSS). 
The HLIS will map 2,000 deg$^2$ of the sky in \textit{Y}, \textit{J}, \textit{H} and \textit{F184} bands with a $5\sigma$ point-source depth of M$_{\mrm{AB},\mathit{J}}$=26.70. 
HLSS will measure slitless galaxy spectra in the same footprint with $7\sigma$ flux sensitivity down to $1.0\,(1.8)\times10^{-16}\rm{erg\,cm^{-2}\,s^{-1}}$ for a point source (extended source, respectively). 
This reference design will be updated and improved with input from the community in order to maximize the scientific return ~\citep[e.g., see ][for a wide survey design that overlaps with the LSST footprint]{WFIRST-LSST}.

This work builds on the kinematic lensing idea as developed in ~\cite{Eric13}. We explore how well the combination of shape measurements from the HLIS and rotation curves extracted from the HLSS data allow for a KL analysis that self-calibrates the intrinsic galaxy shape, thereby significantly reducing the impact of shape noise. We further examine the improved constraining power of the KL measurement with respect to systematic uncertainties and compare our forecasts to the standard \textit{Roman} WL reference survey. As a first step, this work predicts the constraining power on $S_8$-$\Omega_\mrm{m}$ and $w_0$-$w_p$ with shear-shear angular power spectra as the only observable, however we note that an extension to a multi-probe analysis is straightforward and a topic of future work. 

The paper is structured as follows: In Sect.~\ref{sec:KL} we briefly review the KL theory. We then specify our scenario definitions and parameter settings in Sect.~\ref{sec:simulation_settings}. Likelihood modelling, including data vector, covariance matrix and systematics, are described in Sect.~\ref{sec:CosmoLike}. We present our results in Sect.~\ref{sec:results}, and conclude in Sect.~\ref{sec:conclusions}. More details on cosmic shear estimation via KL and effective shape noise evaluation are derived in Appendices~\ref{sec:appd_a} and~\ref{sec:shape_noise_estimation_details}.

\begin{figure*}
    \centering
    \includegraphics[width=\linewidth]{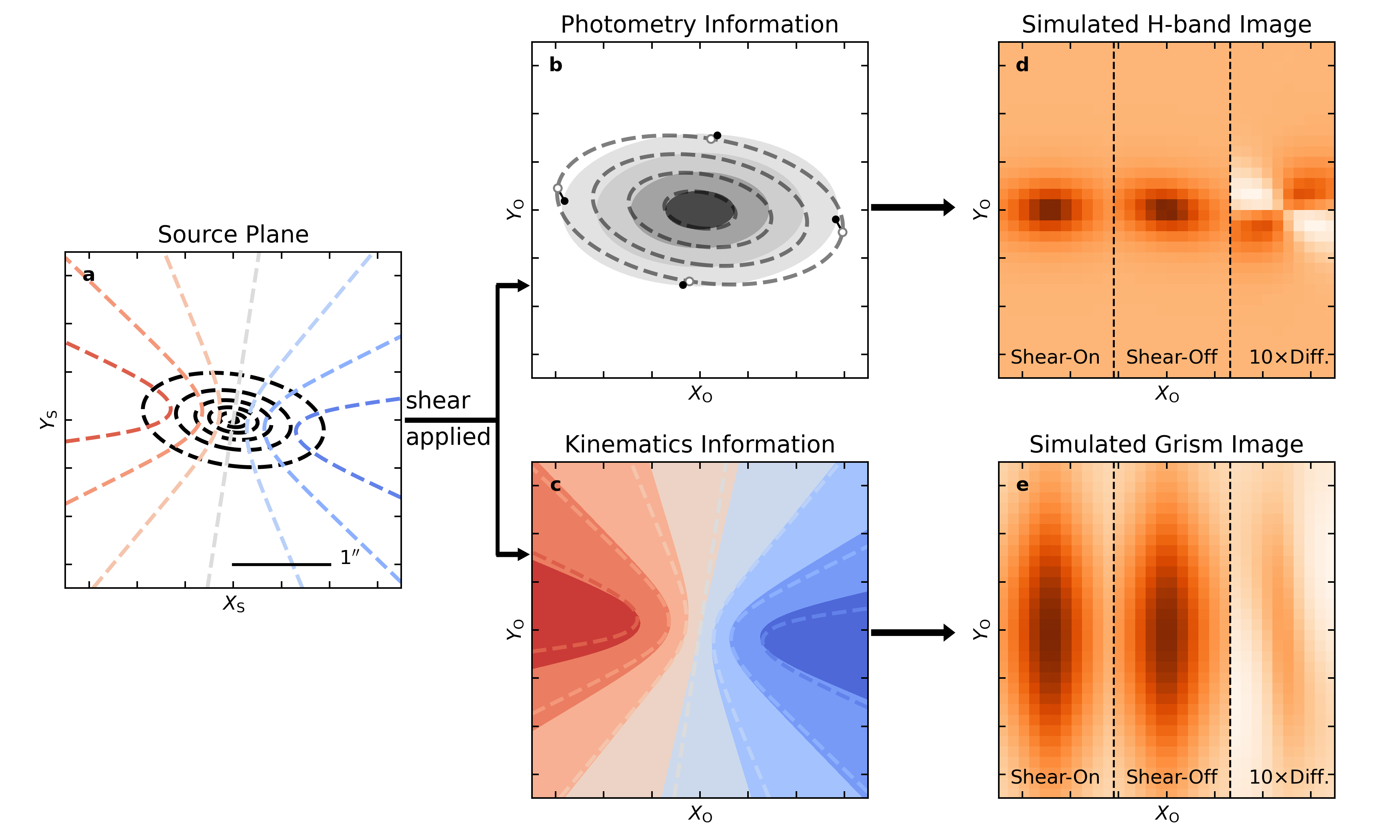}
    \caption{An illustration of the shear effects on photometry image, kinematic map and grism image. 
    \textbf{Panel a:} The projected photometric isophotes (black dashed contour) and LoS velocity map (blue-red dashed contour) as observed in the source plane $X_\mrm{S}$-$Y_\mrm{S}$. 
    \textbf{Panel b:} The projected photometry isophotes observed in the image plane $X_\mrm{O}$-$Y_\mrm{O}$ (grey-black filled contour). As a reference, we also plot the original isophotes before lensing as grey dashed contours. To illustrate the shear distortion better, we take 4 example points on the major and minor axes of the source plane galaxy (the open circles), and show their positions on image plane as filled circles. The displacements are shown in black solid line segments. 
    \textbf{Panel c:} The projected disc velocity map observed in the image plane (blue-red filled contour). We indicate redshift as red and blueshift as blue, relative to the systemic redshift. As a reference, we also plot the original LoS velocity distribution in dashed contour, as in panel a. 
    \textbf{Panel d:} A \textit{Roman} Wide-Field Instrument (WFI)/$H$-band image. From left to right, we show the noise-free simulated data in the image plane (shear-on), the source plane (shear-off), and a 10-times-enhanced difference between the image and source plane.
    \textbf{Panel e:} A \textit{Roman} WFI/Grism image, with dispersion along $Y_\mathrm{O}$ direction. From left to right we show shear-on, shear-off and the 10-times-enhanced difference, similar to panel d. 
    All these panels assume $(\kappa,\gamma)=(0.0, 0.1)$.}
    \label{fig:illustration_src2obs}
\end{figure*}

\section{Kinematic lensing}
\label{sec:KL}
In short, KL is based on the fact that gravitational lensing induced by the large scale structure of the Universe changes the trajectories of photons travelling through space, however their frequencies remain unchanged. The former implies that the shape of a galaxy, aka the photometric image, is distorted when it reaches the observer. The latter implies that the original position of photons can be tightly constrained by the velocity field of a disc galaxy, aka the spectroscopically measured frequency map. 

To quantitatively study the distortion patterns of photometric and kinematic maps under cosmic shear, we firstly define the source plane $X_\mrm{S}$-$Y_\mrm{S}$ as the 2D comoving frame at the position of source galaxy, and the image or observed plane $X_\mrm{O}$-$Y_\mrm{O}$ as the 2D plane at the observer. 
We assume throughout the main sections in this paper that $X_\mrm{O}$-$Y_\mrm{O}$ are aligned with the major-minor axes of the observed galaxy image.
The remapping of the 2D position of photons $\bm \theta$ due to cosmic shear from the source plane to the image plane can be described as $\bm{\theta}_{\mrm{S}}=\mat{A}\bm{\cdot}\bm{\theta}_\mrm{O}$, where the shear transformation matrix is defined as
\begin{equation}
\label{eqn:shear_mat_def_sect2}
    \mat{A} \equiv \frac{\partial\bm{\theta}_\mrm{S}}{\partial\bm{\theta}_\mrm{O}}= (1-\kappa)
    \begin{pmatrix}
    1-g_+ & - g_\times \\
    -g_\times & 1+g_+
    \end{pmatrix}
\end{equation}
and $g= \gamma / (1- \kappa)$ is the reduced shear that can be expressed as a complex number $\bm{g}=g_++ig_\times$ (please see Appendix \ref{sec:appd_a} for a derivation in general coordinates). 

We further illustrate the KL concept in Fig.~\ref{fig:illustration_src2obs}:
\begin{itemize}
    \item Panel a: We show the shape and velocity distribution of the source galaxy without shear distortion.
    \item Panel b: When shear is applied, the distortion pattern on the photometry shape is manifested as a rotation of the image and a change of ellipticity.
    \item Panel c: The distortion pattern of the velocity field is described by a non-axisymmetric transformation.
    \item Panel d: We generate a simulated \textit{Roman} WFI/$H$-band image, with and without shear. The quadrupole structure in the difference map shows how photometry image responds to shear.
    \item Panel e: We generate a simulated \textit{Roman} WFI/Grism image with and without shear. The two distinct patterns in shape (panel b) and velocity distortions (panel c) are combined in the grism image, resulting in the S-shape difference map. The different shear responses in panel d and e give us information on the galaxy intrinsic orientation. This additional information compared to traditional WL, where the intrinsic orientation is treated as an unconstrained noise source, leads to the significant signal-to-noise increase of KL over WL.
\end{itemize}

In the following we detail the methodology of KL shear inference structured into 3 parts: the photometric measurement of the galaxy shapes (Sect.~\ref{sec:KL_morph}), the spectroscopic measurement of the galaxy kinematics (Sect.~\ref{sec:KL_kine}), and the integrated modelling of shapes and kinematics (Sect.~\ref{sec:KL_intg}).

\subsection{Photometric data - galaxy shape}
\label{sec:KL_morph}
To measure galaxy shapes, we use the conventional complex quadratic estimators
\footnote{We are choosing different ellipticity definition than \cite{Eric13} because equation~(\ref{eqn:quadratic_estimators_def}) has simpler transformation rule under shear and is an unbiased estimator of shape~\citep{SS97}}
\begin{equation}
\label{eqn:quadratic_estimators_def}
    \hat{\bm{\epsilon}} = \hat{\epsilon}_1 + i\hat{\epsilon}_2= \frac{I_{11}-I_{22}+2iI_{12}}{I_{11}+I_{22}+2\sqrt{I_{11}I_{22}-I_{12}^2}},
\end{equation}
where
\begin{equation}
\label{eqn:Iij_def}
    I_{ij}=\frac{\int I(\bm{\theta})\theta_i\theta_j\dd^2\bm{\theta}}{\int I(\bm{\theta})\dd^2\bm{\theta}} \,(i,j=1\,\mrm{or}\,2)
\end{equation}
is the 2nd moment of the photometric flux in the galaxy image, with 1 and 2 indicating $X$ and $Y$ components. The original shape before lensing, the so-called intrinsic ellipticity, is denoted as $\hat{\bm{\epsilon}}^\mrm{int}$ and can be expressed through equation~(\ref{eqn:quadratic_estimators_def}) using the intrinsic flux moments of the galaxy.
We can relate the observed ellipticity $e_\mrm{obs}\equiv|\hat{\bm{\epsilon}}|$ to the intrinsic ellipticity $e_\mrm{int}\equiv|\hat{\bm{\epsilon}}^\mrm{int}|$ as (see Appendix~\ref{sec:appd_a})

\begin{equation}
\label{eqn:eobs-eint}
\begin{aligned}
    e_\mrm{int}^2-e_\mrm{obs}^2+2(1-e_\mrm{int}^2)g_+e_\mrm{obs}+(e_\mrm{int}^2e_\mrm{obs}^2-1)g^2=0.
\end{aligned}
\end{equation}
Neglecting terms $\bigo{g^2}$, we derive
\begin{equation}
\label{eqn:shape_result}
    g_+=\frac{e_\mrm{obs}^2-e_\mrm{int}^2}{2e_\mrm{obs}(1-e_\mrm{int}^2)},
\end{equation}
which gives the $g_+$ component of the shear, if both $e_\mrm{int}$ and $e_\mrm{obs}$ are measured.

\begin{figure}
	\includegraphics[width=\linewidth]{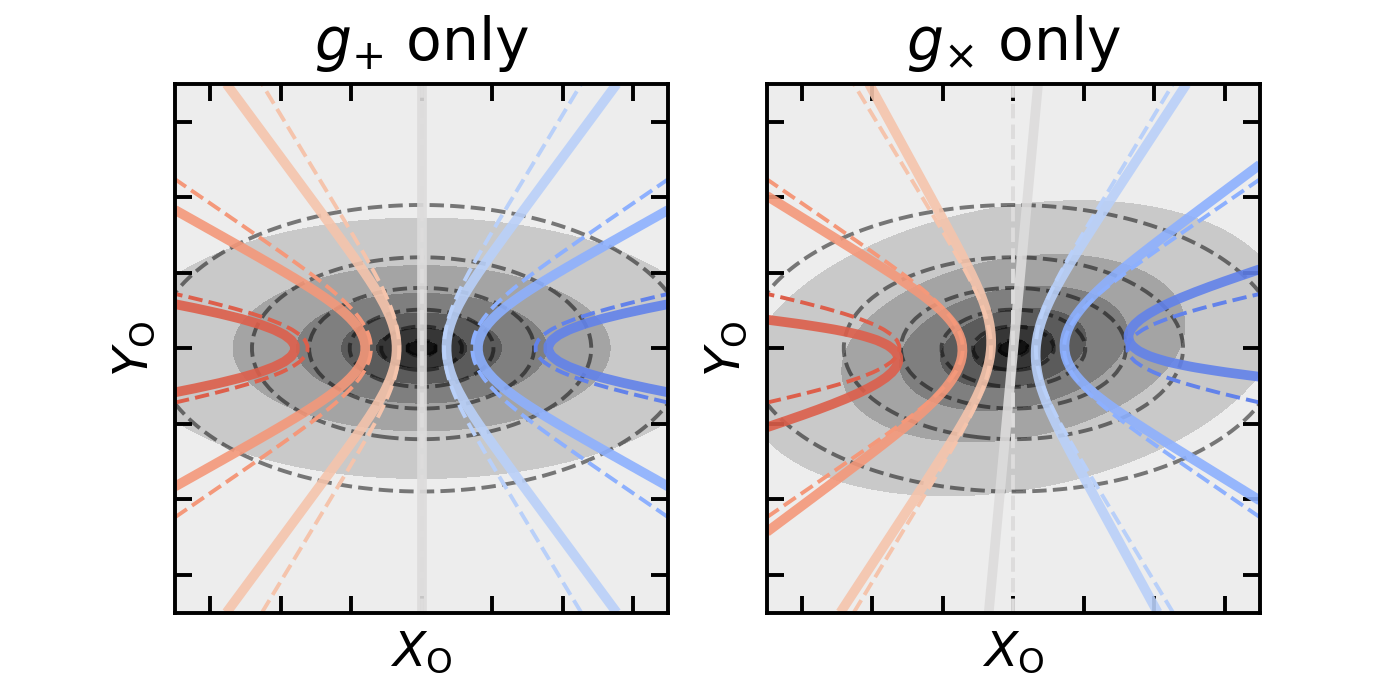}
    \caption{An illustration of the shape and LoS velocity transformation patterns for the different cosmic shear components. We show the impact of $g_+$ on the left and the impact of $g_\times$ on the right. In both the left and the right panels, the observed sheared galaxy shapes are indicated as filled elliptical contours and the original shapes are indicated as unfilled dashed elliptical contours. The observed LoS velocity distributions are shown in solid blue-red contours while the original LoS velocity distributions are plotted as the dashed blue-red contours.
    }
    \label{fig:shear_illus_2}
\end{figure}

We further illustrate the transformation of the photometric components under shear in Fig.~\ref{fig:shear_illus_2}, where the unfilled black dashed contours correspond to the source plane and the black filled contours correspond to the image plane. As shown, $g_+$ changes the ellipticity, while $g_\times$ also rotates the PA. We will further explain in Sect.~\ref{sec:KL_intg}, how $e_\mrm{int}$ can be inferred from  kinematics measurements in combination with the TFR.

\subsection{Spectroscopic data - galaxy kinematics}
\label{sec:KL_kine}

We define the shear transformation of the LoS velocity distribution from source to image plane $v_\mrm{LoS}^\prime$ as
\begin{equation}
\label{eqn:formal_velocity}
    v_\mrm{LoS}^\prime(\bm{\theta}_\mrm{O})=v_\mrm{LoS}(\bm{\theta}_\mrm{S})=v_\mrm{LoS}(\mat{A}\bm{\cdot}\bm{\theta}_\mrm{O}),
\end{equation}
where $v_\mrm{LoS}(\bm{\theta}_\mrm{S})$ is the LoS velocity distribution in the source plane. 

The expression for $v_\mrm{LoS}(\bm{\theta}_\mrm{S})$ in general coordinates is discussed in Appendix~\ref{sec:appd_a}. However, we can gain useful insight by considering the limiting case, where the intrinsic inclination angle $i$ is reasonably large such that the change in PA induced by cosmic shear is much smaller than 1%$e_\mrm{int}\gg g_{+/\times}$ and $\phi^\prime-\phi\ll 1$
\footnote{For nearly face-on source galaxies, a weak shear distortion can change the PA of the photometric major axis significantly, and the expansion of the LoS velocity distribution in the image plane as a Taylor series in $g_{+/\times}$ breaks down.}.
The maximum $v_\mrm{LoS}^\prime$ along the $X_\mrm{O}$-$Y_\mrm{O}$ axis $v_\mrm{major/minor}^\prime$ in the image plane are given by (neglecting terms higher than or equal to $\bigo{g^2}$)
\begin{equation}
\label{eqn:vminor}
    \begin{aligned}
    v^\prime_\mrm{major}&=-v_\mrm{TF}\,\sin{i},\\
    v^\prime_\mrm{minor}&= 
    v_\mrm{TF}\,\sin{i}\,\cos{i}\,(1+\frac{1+e_\mrm{obs}^2}{2e_{\mrm{int}}})g_\times\,,
    \end{aligned}
\end{equation}
where $v_\mrm{TF}$ is the asymptotic circular rotation velocity inferred from the TFR (see Sect.~\ref{sec:KL_intg}). 
Thus $v_\mrm{major}^\prime$ is a good proxy of $\sin{i}$, and $v_\mrm{minor}^\prime$ is directly probing $g_{\times}$, which are also consistent with Fig.~\ref{fig:shear_illus_2} (the solid and dashed blue-red contours): The LoS velocity along $X_\mrm{O}$ axis is agnostic against $g_+$ and $g_\times$, but the velocity along $Y_\mrm{O}$ is sensitive to $g_\times$ due to its non-axisymmetric impact.

We note that \cite{Miller13} suggest that for typical disc galaxies, the rotational velocity converges to the asymptotic rotational velocity at 2.2 times the galaxy scale radius ($r_{2.2}$). Our method requires that we selected galaxies where velocity measurements beyond $r_{2.2}$ can be made with sufficient SNR.

Kinematic measurements based on slitless spectroscopy, such as the \rst\ grism, are challenging. The Doppler shift information is entangled with the galaxy spectrum and the photometric profile as shown in panel e in Fig.~\ref{fig:illustration_src2obs}. 
Encouragingly, ~\cite{OC20} have successfully measured galaxy kinematics at a precision of $\pm$40 km$\,$s$^{-1}$ using the \textit{HST} grism (the \rst\ grism will have a factor of 3--4 higher resolution).  

\subsection{Integrated Modelling}
\label{sec:KL_intg}

In order to break the degeneracy between intrinsic shape and cosmic shear, we first isolate $v_\mrm{TF}$ from $v_\mrm{major}^\prime$.
The TFR allows us to express ~\citep{R11}
\begin{equation}
\label{eqn:TFR}
    \mrm{log}_\mathrm{10}(v_{\mrm{TF}})=a + b\, (M_\mrm{B}-M_\mrm{p})\,, 
\end{equation}
where $M_\mrm{B}$ is the absolute magnitude and $M_\mrm{p}$ is the pivot value, which is the weighted sample mean of $M_\mrm{B}$. $a$ and $b$ are the zero-point and the slope of the scaling relation. 
The intrinsic scatter of the TFR calibrated in $i$-band is 0.049 dex and even smaller scatter can be achieved with more colour information~\citep{R11}. 

Once $i$ is inferred per galaxy from the TFR and velocity measurement $v_\mrm{major}^\prime$, the intrinsic ellipticity $e_\mrm{int}$ is derived from a simple geometric relation
\begin{equation}
\label{eqn:eint_geo}
    e_{\mrm{int}}=\frac{1-\sqrt{1-(1-q_z^2)\sin{^2i}}}{1+\sqrt{1-(1-q_z^2)\sin{^2i}}}\,,
\end{equation}
where $q_z$ is the ratio of scale height to scale radius of the disc galaxy.  
Typical $q_z$ is around 0.25, and is larger for galaxies at higher redshift, since disc galaxies are more turbulent at earlier time~\citep{Ubler:2017pik}. 
When estimating shear for individual galaxies, $q_z$ is not known perfectly and should be treated as a nuisance parameter in the KL shear inference pipeline.

Combining equation~(\ref{eqn:shape_result}), (\ref{eqn:vminor}), (\ref{eqn:TFR}) and (\ref{eqn:eint_geo}), we derive
\begin{equation}
\label{eqn:reduced_shears}
    \begin{aligned}
    \hat{g}_+ &= \frac{e_\mrm{obs}^2-e_\mrm{int}^2}{2e_\mrm{obs}^2(1-e_\mrm{int}^2)},\\
    \hat{g}_\times&=|\frac{v_\mrm{minor}^\prime}{v_\mrm{major}^\prime}|\frac{2e_\mrm{int}}{\cos{i}\,(2e_\mrm{int}+1+e_\mrm{obs}^2)}\,,
    \end{aligned}
\end{equation}
which allows us to estimate both shear components given the measured ellipticity $e_\mrm{obs}$, the measured velocities $v_\mrm{minor}^\prime$ and $v_\mrm{major}^\prime$, the inclination angle $i$ as computed in equation~(\ref{eqn:vminor}) and (\ref{eqn:TFR}), and the inferred intrinsic ellipticity $e_\mrm{int}$ as computed in equation~(\ref{eqn:eint_geo}).  

In practice, instead of directly using the estimators described in equation~(\ref{eqn:reduced_shears}), we recommend building a generative full-forward model to predict the sheared galaxy photometry and grism images at pixel level. Such a generative model has to account for realistic systematics, e.g. point-spread function (PSF), galaxy morphology and aspect ratio $q_z$, and uncertainties in the TFR. 

\section{Galaxy sample characteristics}
\label{sec:simulation_settings}

Based on the reference survey design further described in \cite{WFIRST_MultiProbe} we compare two science cases: 
\begin{itemize}
    \item \textbf{Traditional Weak Lensing (WL)}: cosmological information from shear-shear two-point statistics using the reference survey HLIS imaging dataset only. 
    \item \textbf{Kinematic Lensing (KL)}: cosmological information from shear-shear two-point statistics using both photometric and spectroscopic data. This effectively limits our galaxy sample to that of the HLSS dataset and even adds some additional constraints.
\end{itemize}

\subsection{The WL galaxy sample}
\label{sec:sampleselection_WL}

The WL sample assumed here is the same as for the reference \rst\ HLIS ~\citep{WFIRST_MultiProbe}. Based on the CANDELS~\citep{CANDELS} catalogue, realistic sources are selected following \cite{hemmati2019}. The reference HLIS is expected to have a $5\sigma$ depth of 26.54 in \textit{H}-band; given that the depth of the CANDELS (the GOODS-South Field, which is one of the two CANDELS Deep Program fields) reaches a $5\sigma$ point source limit of 27.6 mag in \textit{F160W}, we can assume that the catalogue is complete. 
The specific selection criteria for our WL sample are
\begin{itemize}
    \item \textit{J}+\textit{H} band combined SNR $>18$
    \item ellipticity measurement error $\sigma(\bm{\epsilon})<0.2$\footnote{In this and only in this criterion, we take $\bm{\epsilon}=(I_{11}-I_{22}+2iI_{12})/(I_{11}+I_{22})$, as in~\cite{BJ02}. $\sigma(\bm{\epsilon})$ is defined as rms per $\bm{\epsilon}$ component, i.e. the observed shape uncertainty due to image noise, PSF and pixelization. Note that this is different from shape noise, which captures the uncertainty from random intrinsic inclination angle.}
    \item spatial resolution factor $R>0.4$\footnote{$R\equiv(1+\mrm{EE50}_\mrm{PSF}^2/r_\mrm{eff,gal}^2)^{-1}$, where $\mrm{EE50}$ is the 50 per cent encircled energy radius and $r_\mrm{eff,gal}$ the galaxy effective radius~\citep{ETC}.}
\end{itemize}
The resulting average number density is 51 arcmin$^{-2}$, and the redshift distributions including the split into 10 tomographic bins of equal number of galaxies are shown in Fig.~\ref{fig:zdist} (green lines).

\subsection{The KL galaxy sample}
\label{sec:sampleselection_KL}
We use the COSMOS Mock Catalogue~\citep[CMC, ][]{CMC} to infer the number density and redshift distribution of source galaxies that qualify for a KL measurement with the \rst. Based on the observed COSMOS catalogue, the CMC fits a template for each galaxy using the observed spectral energy distribution and predicts its photometry and emission line strength ($\mrm{H}\,\alpha$, [O\,\textsc{iii}], etc.). The emission line fluxes are calculated from empirical relations between emission line fluxes and dust-corrected UV star formation rates and are further calibrated with a spectroscopy sample~\citep[see Sect.~2.2.3 in ][and reference therein for more details.]{CMC}\par

We select galaxies with high SNR to ensure robust rotation velocity measurements, and also require that our galaxies have sufficient angular size such that the rotation velocity converges to $v_{\rm{TF}}$ at 2.2 times the scale radii. 
We impose the following criteria
\begin{itemize}
    \item At least one of the emission lines ($\rm{H}\,\mrm{\alpha}, \rm{H}\,\mrm{\beta}$, [O\,\textsc{iii}]) tracing the disc kinematic of galaxy is resolved within the grism's spectral range. %(between $1\mu m \sim2\mu m$)
    \item Emission line flux reach $7\sigma$ point-source detection limit, which is $10^{-16}\ \rm{erg\,s^{-1}\,cm^{-2}}$
    \item Half-light radius $\geq$ $0.1\arcsec$. As a comparison, the \textit{Roman Space Telescope} PSF is around $0.1\arcsec$
    \item \textit{z}-band magnitude $\leq$24.5
\end{itemize}

\begin{figure}
	\includegraphics[width=\columnwidth]{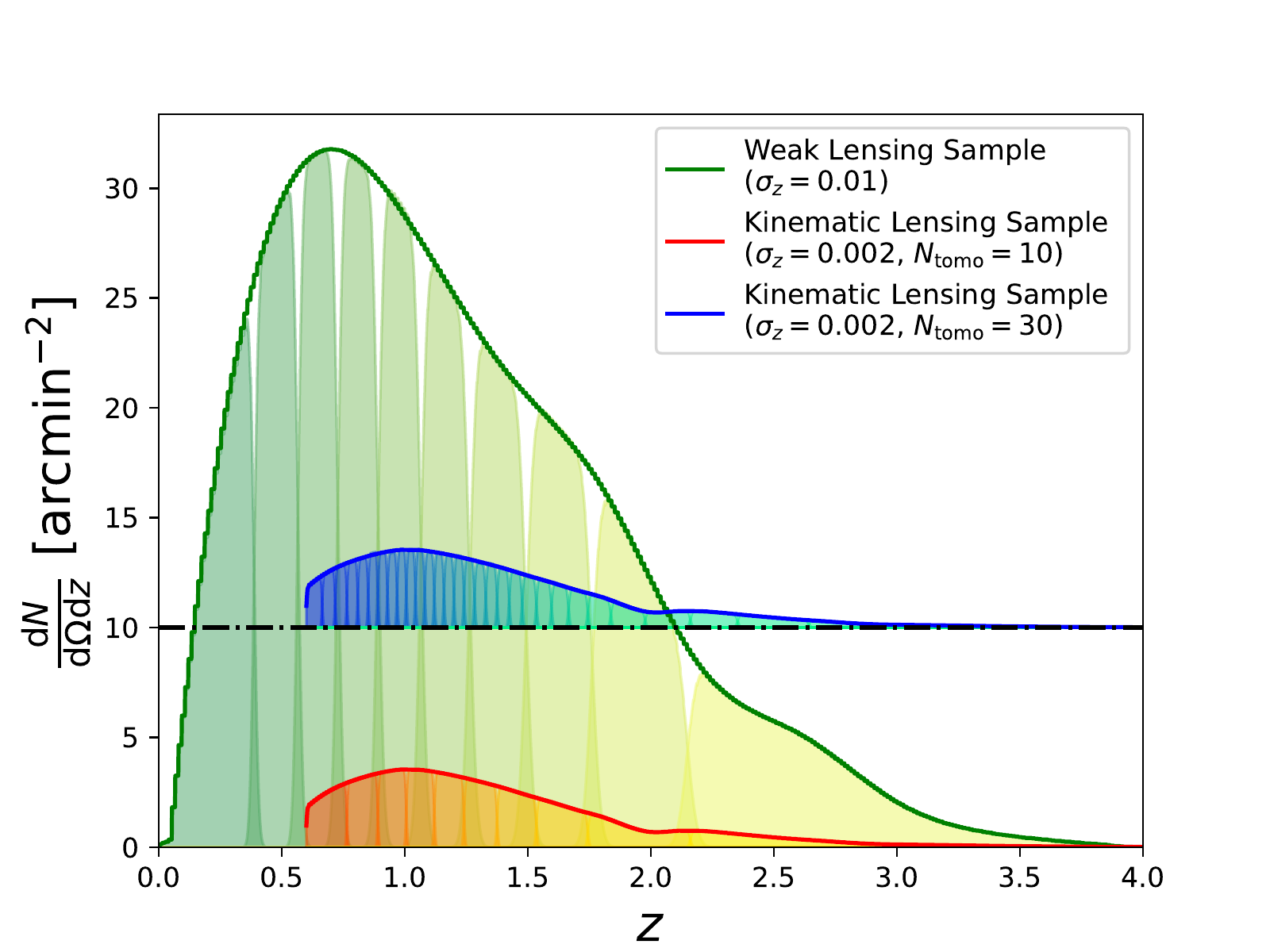}
    \caption{The redshift distribution of source galaxies per tomographic bin in traditional weak lensing method (the green solid lines, according to the baseline \textit{Roman Space Telescope} HLIS), kinematic lensing method with 10 tomography bins (the red solid lines, according to baseline \textit{Roman Space Telescope} HLSS), and kinematic lensing method with 30 tomography bins (the blue solid lines, offset by 10 arcmin$^{-2}$ for better visualisation).}
    \label{fig:zdist}
\end{figure}

The resulting source galaxy density is 8.13 arcmin$^{-2}$. We further require that the galaxy sample also satisfies the criteria in Sect.~\ref{sec:sampleselection_WL}, however the HLIS selection criteria don't lead to a further reduction of the KL sample.
We further multiply this number by a factor of 0.5, in order to account for observational inefficiencies, arriving at 4 arcmin$^{-2}$ as the number entering our likelihood analysis. The redshift distributions are depicted in Fig.~\ref{fig:zdist} (the red/blue lines for $N_{\mrm{tomo}}=10/30$ scenario). 

We also compare this galaxy number density with other estimations in the literature. 
The EL-COSMOS catalogue~\citep{sti20} is based on the updated COSMOS2015 catalogue~\citep{COSMOS2015} and more physically motivated emission line modelling. 
Applying the same cuts on the emission line SNR\footnote{The EL-COSMOS catalogue does not have half-light radius attribute, so we neglect the criterion of hlr $\geq 0.1\arcsec$. However the resulting bias is negligible since almost all the galaxies satisfy this criterion.}, we obtain $n_{\mathrm{gal}}\approx 12\,\mathrm{arcmin}^{-2}$, before accounting any observational inefficiency. 
\cite{zbw19}, on the other hand, uses semi-analytic modelling to predict the H$\,\alpha$ emitter density for $Euclid$ and $Roman$. 
For a combined $z$=0.5--3.0 [O\,\textsc{III}] and H$\,\mathrm{\alpha}$ sample with high-redshift dust extinction model, the galaxy surface number density is around 4.14 arcmin$^{-2}$ without any observational inefficiency considered.

This number is somewhat different compared to the estimation from the EL-COSMOS catalogue, especially given both of them are calibrated with the \Ha\ luminosity functions from HiZELS~\citep{Sobral13}. 
However, we note that the \Ha\ luminosity functions predictions in \cite{sti20} and \cite{zbw19} do not agree perfectly with each other, and \Ha\ emitters at $z\approx 1$, where our $\dd N/\dd \Omega\dd z$ peaks, are not included in the HiZELS luminosity functions. 
Thus we think the difference in galaxy number density is a reasonable manifestation of the significant differences in these two methods. 
Our number (8 arcmin$^{-2}$ without accounting for observational inefficiencies) falls somewhere between these two results.
We note that the same trend is also found in \cite{FC20}, where the authors show that the number density of both the \Ha\ and the \OIII\ samples in \cite{zbw19} are somewhat lower than the predictions from HiZELS~\citep{ksm15}.

Another estimate is obtained from table~2-2 in \cite{WFIRST_AFTA}, which predicts a sample of 17.8 million emission line galaxies (\Ha\ and \OIII) in a 2,200 deg$^2$ survey, or 2.25 arcmin$^{-2}$. This number however includes observational inefficiencies, and is based on the old HLSS design, where only emission line galaxies within $1.06\leq z\leq 2.77$ are observable. 
Since in our reference HLSS sample, \Ha+\OIII\ emitters within $1.06\leq z\leq 2.77$ take up 61.91 per cent of the total sample, an approximate scaling of the prediction in ~\cite{WFIRST_AFTA} gives 3.63 arcmin$^{-2}$, which is very similar to the number assumed in this paper.

We conclude that our assumed number density of galaxies is in line with other estimates in the literature even though these are based on different techniques such as semi-analytic modelling and analytic models based on an assumed luminosity function. 
We also consider the reduction of the number density estimate by a factor of 2 to account for observational inefficiencies as conservative. 

%%% Figure 2: the ztrue distribution for each tomo bin
%%% =================================================

\subsection{Shape Noise}
\label{sec:shape_noise}

For our WL sample we assume a shape noise level of 0.26 per component and $\sigma_{\bm{\epsilon}}^\mrm{WL}=0.37$ in quadrature, which is standard across the community~\citep[see e.g.,][]{Chang13}.
The shape noise of our KL sample is $\sigma_{\bm{\epsilon}}^\mrm{KL}=0.035$ in quadrature, which we explain further below.

From the source selection criteria in Sect.~\ref{sec:sampleselection_KL} we know that the faintest sources in the KL sample have $r$-band magnitudes around 26, and 95.7 per cent of the sample has $M_r<25$. 
The LSST Year 10 $5\sigma$ $r$-band limit is 27.5, which means most of our KL sample have $r$-band SNR higher than 50. 
Also, according to the CMC catalogue, 97.1 (69.3) per cent of the sample have hlr larger than 1 (2) times the typical PSF hlr of \rst, and 39.3 per cent of the sources have hlr larger than the LSST $r$-band PSF. 

Given the high S/N we expect the dominant contribution to KL shape noise to come from velocity measurement error and the intrinsic scatter of TFR. To estimate this effective shape noise per galaxy, we build an error model for the KL observables and numerically calculate the maximum likelihood estimate (MLE). 

The exact details of the procedure can be found in Appendix.~\ref{sec:shape_noise_estimation_details}. In short, we create random samples of KL model parameters $\bm{X}_\mathrm{model}$=($e_\mrm{int}$, $\phi$, $\bm{g}$, $v_\mrm{TF}$) where $\phi$ is the intrinsic PA of the galaxy image in source plane. For each realization, we calculate the data vector of observables $\bm{Y}$=($\hat{\bm{\epsilon}}$, $v_\mrm{major}^\prime$, $v_\mrm{minor}^\prime$, $M_\mrm{B}$) according to equation~(\ref{eqn:estimators_image_to_source}), (\ref{eqn:galaxy_frame_source_plane_velocity}) and (\ref{eqn:TFR}). 
Realistic measurement errors are then added to those observables. These errors determine the covariance matrix for the multivariate Gaussian log-likelihood function $\mrm{ln}(p(\bm{Y}|\bm{X}_\mathrm{model}))$. 
We then call optimization routines to get MLE solution of $\bm{g}_{i}$ for each realization. 
The shape noise is finally computed as the $\sigma^{-2}(g_{+/\times})$ weighted standard deviation among the random sample. 

\begin{figure}
    \centering
    \includegraphics[width=\linewidth]{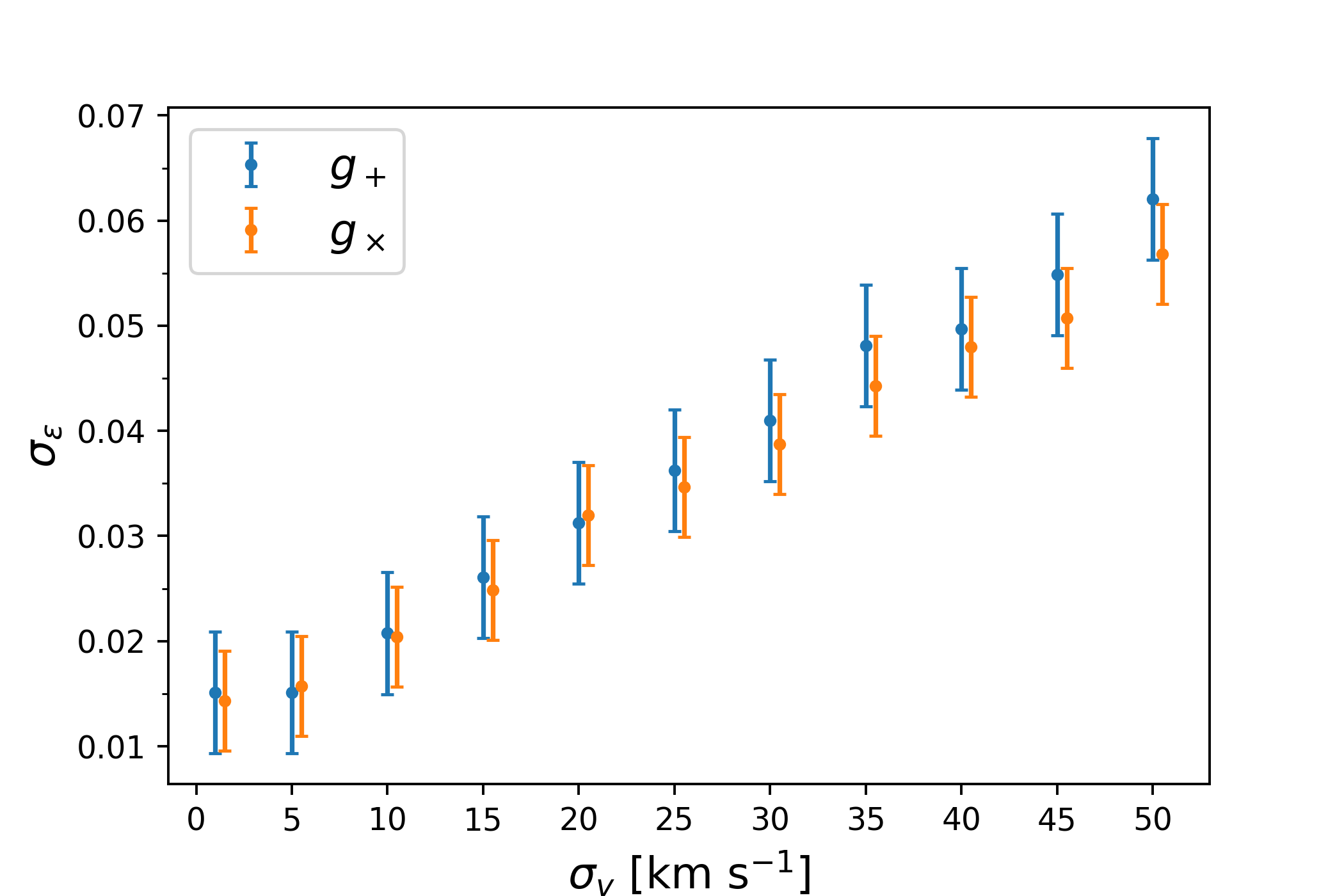}
    \caption{Here we show the effective shape noise $\sigma_{\bm{\epsilon}}$ as a function of velocity measurement errors $\sigma_v$. Blue for the $g_+$ component and orange for the $g_\times$ component. The shape noise is calculated as weighted standard deviation of the random sample, while the error bars on shape noise are estimated via a bootstrap method. 
    %Data points are measured at $\sigma_v$=1, 5, 10, 20, 30, 40 and 50 km$\,$s$^{-1}$. 
    The $g_\times$ data points are slightly offset for better visualization. }
    \label{fig:shape_noise_velocity_error}
\end{figure}

We show the shape noise as a function of velocity measurement uncertainty $\sigma_v$ in Fig.~\ref{fig:shape_noise_velocity_error}, with the blue and orange points for $g_+$ and $g_\times$. 
As we can see, for $\sigma_v>10$ km$\,$s$^{-1}$, the shape noise is linearly increasing with $\sigma_v$. 
For $\sigma_v<10$ km$\,$s$^{-1}$, the shape noise is gradually being saturated by the intrinsic scatter of TFR. 

We note that a similar galaxy kinematic measurement on grism data has already been carried out recently by~\cite{OC20}, although without accounting for cosmic shear distortions. 
The authors built a forward modelling pipeline and applied it to the \textit{Hubble Space Telescope} based datasets, 3D-HST and GLASS. 
\cite{OC20} table~2 shows the kinematic measurements of 4 galaxies. 
The rotation velocity uncertainties are 30--40 km$\,$s$^{-1}$ for single roll exposures and $15$ km$\,$s$^{-1}$ if more roll angles are available. 
The higher spectral resolution of \rst\ grism, its grism observing strategy, and the bright nature of our KL sample, should enable typical KL velocity measurement errors to be 15 km$\,$s$^{-1}$ or better. 
Using Fig.~\ref{fig:shape_noise_velocity_error}, this translates to a shape noise of $\sim 0.025$ per component, or 0.035 in quadrature. 

Similar kinematic lensing measurements have also been conducted by \cite{GTF20,GTF21,DiGiorgio21}. 
Those works are different from ours in both theoretical methodology and the instrument used for observations: they use the lens galaxies position rather than the TFR to provide extra information on $g_+$, and use ground-based integral field units (IFU) rather than space-based grism.
Thus their sample is focused on low-$z$ bright and well-resolved disc galaxies in galaxy-galaxy lensing systems, where the tangential shear contributed by the lens galaxies dominate over the cosmic shear effect. 
The resulting sample contains 18 galaxies with an averaged tangential shear of $\langle\gamma_\mrm{t}\rangle=0.020\pm 0.008$. 
The effective shape noise per galaxy originated from velocity measurement uncertainty ($\sigma_\gamma$ in their table~3) is 0.013. 

Despite the very different nature between their sample and ours, the two samples have comparable photometric SNR and spatial resolution factor ($R\approx 0.8$). 
The difference in shape noise could come from velocity resolving power $\mathcal{R}_\mrm{kin}=\frac{c}{(1+z)\mathcal{R}}$, the velocity dispersion per pixel. 
For the sample in~\cite{GTF20}, $\mathcal{R}_\mrm{kin}\approx 90$, while for \rst\ grism $170<\mathcal{R}_\mrm{kin}<330$. 
Assuming that shape noise scales as $\mathcal{R}_\mrm{kin}$, then we expect the shape noise of our KL sample is 2--4 times larger, which is 0.025--0.051 and is consistent with our $\sigma_{\bm{\epsilon}}^\mrm{KL}$. 

\cite{DiGiorgio21} also studied shape noise as a function of inclination angle, image PA uncertainty and velocity error. 
For $\sigma_v$=10--20 km$\,$s$^{-1}$, $\sigma_\mrm{PA}$=1--6 deg and $i<45$ deg, the shape noise ranges from 0.015 to 0.08, which is of the same order as our result. 

A comprehensive kinematic lensing shape noise study, highly relevant to our work, is presented in~\cite{WS19}, which considers TFR as an ingredient to break intrinsic shape-shear degeneracy. 
The authors utilized Fisher matrix analysis to map out shape noise as a function of various variables, including inclination angle, TFR scatter, orthogonal-slits measurement v.s. IFU velocity map, velocity measurement uncertainty, etc. 
Assuming that the shape noise measured at $i=\pi/4$ is a good representation of the ensemble average, we expect to see a shape noise of 0.036--0.054 for orthogonal-slit measurement with $\sigma_v\approx 20$ km$\,$s$^{-1}$ and a fiducial $g\approx 0.05$.  

A detailed study of KL shape noise is presented in \cite{SKH+22}, where the authors build a slit-spectrum-based KL shear inference pipeline and study shape noise as a function of SNR, intrinsic ellipticity, spectral resolution, and TFR uncertainty using Keck/DEIMOS-like simulated data. Keck/DEIMOS has higher spectral resolution but worse PSF control compared to \textit{Roman} grism. \cite{SKH+22} find the average shape noise of a galaxy sample to be $0.02$--$0.06$.

Given the differences between the KL implementation of this paper compared to the aforementioned, which range from instrumentation to sample selection to methodology, we note that these comparisons serve as sanity checks rather than direct comparisons. 
A more realistic shape noise calculation requires a more sophisticated pipeline and has to deal with complexities like non-axisymmetric disc structure, magnification-shear degeneration and other systematics, which are beyond the scope of this work. 
We conclude that $\sigma_{\bm{\epsilon}}^\mrm{KL}=0.035$ is a good assumption in order to explore the science outcome of a KL survey conducted with \rst, and to explore multi-probe and cross-survey synergies using KL.

\begin{table*}
	\centering
	\begin{tabular}{l|cccc}
	\toprule
	Cosmology Parameters & \multicolumn{2}{c}{Fiducial} & \multicolumn{2}{c}{Prior}\\
	\midrule
	$\Omega_\mrm{m}$&\multicolumn{2}{c}{0.3156}&\multicolumn{2}{c}{flat[0.095, 0.585]}\\
	$\sigma_8$&\multicolumn{2}{c}{0.831}&\multicolumn{2}{c}{flat[0.5, 1.1]}\\
	$n_s$&\multicolumn{2}{c}{0.9645}&\multicolumn{2}{c}{flat[0.84, 1.06]}\\
	$\Omega_\mrm{b}$&\multicolumn{2}{c}{0.0491685}&\multicolumn{2}{c}{flat[0.005, 0.095]}\\
	$h_0$&\multicolumn{2}{c}{0.6727}&\multicolumn{2}{c}{flat[0.4, 0.9]}\\
	$w_0$($\Lambda$CDM)&\multicolumn{2}{c}{-1}&\multicolumn{2}{c}{flat[-2.1, 0]}\\
	$w_a$($\Lambda$CDM)&\multicolumn{2}{c}{0}&\multicolumn{2}{c}{flat[-2.6, 2.6]}\\
	$w_0$($w$CDM 1)&\multicolumn{2}{c}{-0.289}&\multicolumn{2}{c}{flat[-1.389, 0.711]}\\
	$w_a$($w$CDM 1)&\multicolumn{2}{c}{-2.21}&\multicolumn{2}{c}{flat[-4.18, 0.39]}\\
	$w_0$($w$CDM 2)&\multicolumn{2}{c}{-1.249}&\multicolumn{2}{c}{flat[-2.349, -0.249]}\\
	$w_a$($w$CDM 2)&\multicolumn{2}{c}{0.59}&\multicolumn{2}{c}{flat[-2.01, 3.19]}\\
	\toprule
	\thead{Survey\\Parameters}&\multicolumn{2}{c}{WL}&\multicolumn{2}{c}{KL}\\
	\midrule
	$\Omega_{\mrm{s}}$ [deg$^2$]&\multicolumn{2}{c}{2,000}&\multicolumn{2}{c}{2,000}\\
	$n_{\mrm{src}}$ [arcmin$^{-2}$]&\multicolumn{2}{c}{51}&\multicolumn{2}{c}{4}\\
	$\sigma_{\bm{\epsilon}}$&\multicolumn{2}{c}{0.37}&\multicolumn{2}{c}{0.035\,($N_{\mrm{tomo}}=10$)/0.056\,($N_{\mrm{tomo}}=30$)}\\
	\toprule
	\thead{Systematic\\Parameters} & \multicolumn{2}{c}{WL} & \multicolumn{2}{c}{KL}\\
	\hline
	{} & Fiducial & Prior & Fiducial & Prior \\ 
	\midrule
	$\Delta_{\mrm{z,\,src}}^{i}$&0.0&$\Ndist{0}{2}{-3}$&0.0&$\Ndist{0}{4}{-4}$\\
	$\sigma_{\mrm{z,\,src}}^{i}$&0.01&$\Ndist{0.01}{2}{-3}$&0.002&$\Ndist{0.002}{4}{-4}$\\
	$m^{i}$&0.0&$\Ndist{0}{2}{-3}$&0.0&$\Ndist{0}{4}{-4}$\\
	\hline
	$A_{\mrm{IA}}$&5.92&$\mathcal{N}$(5.92,\,3.0)&-&-\\
	$\beta_{\mrm{IA}}$&1.1&$\mathcal{N}$(1.1,\,1.2)&-&-\\
	$\eta_{\mrm{IA}}$&-0.47&$\mathcal{N}$(-0.47,\,3.8)&-&-\\
	$\eta_{\mrm{IA}}^{\mrm{high}\textit{-}z}$&0.0&$\mathcal{N}$(0.0,\,2.0)&-&-\\
	$Q_1$&0.0&$\mathcal{N}$(0.0,\,16.0)&0.0&$\mathcal{N}$(0.0,\,16.0)\\
	$Q_2$&0.0&$\mathcal{N}$(0.0,\,2.0)&0.0&$\mathcal{N}$(0.0,\,2.0)\\
	$Q_3$&0.0&$\mathcal{N}$(0.0,\,0.8)&0.0&$\mathcal{N}$(0.0,\,0.8)\\
	\bottomrule
	\end{tabular}
	\caption{The parameter settings for different scenarios in \CL simulations. We express a flat prior as flat[min,max] where min and max are the boundary of the uniform distribution. We express a Gaussian prior as $\mathcal{N}(\mu,\sigma)$ where $\mathcal{N}$ means normal distribution, $\mu$ is the mean and $\sigma$ is the standard deviation. We fill the IA parameters ($A_{\mrm{IA}}$, $\beta_{\mrm{IA}}$, $\eta_{\mrm{IA}}$, $\eta_{\mrm{IA}}^{\mrm{high}\textit{-}z}$) in KL scenario with `-' to indicate that the IA mitigation method doesn't apply to the KL sample. Note that we had to increase the shape noise for the 30 tomography bins scenario due to numerical stability problems of the covariance matrix.
	We also adopt a conservative redshift uncertainty for the KL sample in this work.
	}
	\label{tab:params_table}
\end{table*}

\section{Simulated Likelihood Analysis}
\label{sec:CosmoLike}

Our KL \CL~\citep{Krause_Tim17} implementation for the simulated analyses follows the widely used Bayesian approach to infer the posterior of cosmological parameters $\bm{p}_{\mrm{co}}$ and nuisance parameters $\bm{p}_{\mrm{nu}}$ from the observables, which are treated as data vector $\bm{D}$, and prior information $p(\bm{p}_{\mrm{co}},\,\bm{p}_{\mrm{nu}}|I)$\footnote{Here we use $I$ to discriminate different external datasets as priors, although we only adopt flat priors on cosmological parameters in this work.}.
We use \textsc{emcee}~\citep{emcee,goodman2010} to sample the posterior
\begin{equation}
\label{eqn:post}
\begin{split}
    p(\bm{p}_{\mrm{co}},\,\bm{p}_{\mrm{nu}}|\bm{D},\,I)&\propto L(\bm{D}|\bm{p}_{\mrm{co}},\,\bm{p}_{\mrm{nu}} ) \times p(\bm{p}_{\mrm{co}},\,\bm{p}_{\mrm{nu}}|I)\\
    &\propto\mrm{exp}\left(-\frac{1}{2} \chi^2 (\bm{p}_\mrm{co},\,\bm{p}_\mrm{nu})\right)\times 
    p(\bm{p}_\mrm{co},\,\bm{p}_\mrm{nu}|I).
\end{split}
\end{equation}
We assume that the likelihood $L(\bm{D}|\bm{p}_{\mrm{co}},\,\bm{p}_{\mrm{nu}} )$ in equation~(\ref{eqn:post}) is well-approximated by a multivariate Gaussian \citep{likelihood}.

We define  
\begin{equation}
\begin{aligned}
    \chi^2(&\bm{p}_{\mrm{co}},\,\bm{p}_{\mrm{nu}})=\\&(\bm{D}-\bm{M}(\bm{p}_{\mrm{co}},\,\bm{p}_{\mrm{nu}}))^{\rm{T}}\bm{\cdot}\mat{C}^{-1}\bm{\cdot}(\bm{D}-\bm{M}(\bm{p}_{\mrm{co}},\,\bm{p}_{\mrm{nu}})), 
\end{aligned}
\end{equation}
where $\bm{M}$ is the model vector evaluated at $(\bm{p}_{\mrm{co}},\,\bm{p}_{\mrm{nu}})$ and $\mat{C}$ the covariance matrix calculated at the fiducial parameter values. We further assume that the covariance matrix $\mat{C}$ of the multivariate Gaussian is constant within the range of parameters considered and do not adopt an iterative inference approach or a full cosmology dependent covariance ~\citep{ESH_09} approach in this work. 

\subsection{Data Vector}
We only consider the shear-shear two point statistics as our observable.
In practice, results based on $\bm{\xi}^{ij}_\pm(\theta)$ and $\bm{C}_{\kappa\kappa}^{ij}(\ell)$ do not necessarily agree with each other perfectly~\citep[e.g. see~][]{HSC_Y1_2PCF} due to the weighing of the different scales, however we assume that this difference is negligible for the forecasting purpose of this paper. 
Due to the convenience of Fourier space modelling, we choose the angular shear-shear power spectrum as our observable, which in the Limber approximation reads:
\begin{equation}
    \bm{C}_{\kappa\kappa}^{ij}(\ell)=\int \dd\chi \frac{q_\kappa^{i}(\chi)q_\kappa^{j}(\chi)}{\chi^2}P_{\delta\delta}(\ell/f_{K}(\chi),\,z(\chi)),
\end{equation}
where
\begin{equation}
    q_\kappa^{i}(\chi)=\frac{3H_0^2\Omega_\mrm{m}}{2c^2}\frac{\chi}{a(\chi)}\int_\chi^{\chi_h}\dd \chi'\frac{n_\mrm{src}^{i}(z(\chi'))}{\bar{n}_\mrm{src}^{i}}\frac{\dd z}{\dd \chi'}\frac{f_K(\chi'-\chi)}{f_K(\chi')}
\end{equation}
is the lens efficiency for the $i$-th bin, and $n_\mrm{src}^{i}(z)$ represents the redshift distribution of source galaxies in the $i$-th bin (c.f. Fig.~\ref{fig:zdist}), with $\bar{n}_\mrm{src}^{i}$ the 2D angular galaxy density within the same redshift bin. 
Here, $\chi$ is the comoving distance, $f_K(\chi)$ is the comoving angular diameter distance, $\mathbf{\ell}$ the wave vector and $P_{\delta\delta}(\ell/f_K(\chi),z(\chi))$ is the 3D matter power spectrum.

\subsection{Covariance Matrix}
Consider the covariance matrix between $\bm{C}_{\kappa\kappa}^{ij}(\ell_1)$ and $\bm{C}_{\kappa\kappa}^{kl}(\ell_2)$
\begin{equation}
\begin{aligned}
    \mat{C}&\equiv \langle \Delta \bm{C}_{\kappa\kappa}^{ij}(\ell_1) \Delta \bm{C}_{\kappa\kappa}^{kl}(\ell_2)\rangle\\
    &=\langle (\bm{C}_{\kappa\kappa}^{ij}(\ell) - \langle{\bm{C}}_{\kappa\kappa}^{ij}(\ell)\rangle)(\bm{C}_{\kappa\kappa}^{kl}(\ell_2) - \langle{\bm{C}}_{\kappa\kappa}^{kl}(\ell_2)\rangle)\rangle,
\end{aligned}
\end{equation}
where $\langle{\bm{C}}_{\kappa\kappa}^{ij}(\ell)\rangle$ is the mean angular power spectrum among an ensemble of random realizations.
The covariance matrix \mat{C} is modelled as the sum of a Gaussian part $\mat{C}_{\rm{G}}$ and a Non-Gaussian part $\mat{C}_{\rm{NG}}$. The Gaussian part is~\citep{HB04}
\begin{equation}
\label{eqn:C_G}
\begin{aligned}
    \mat{C}_{\mrm{G}} (\bm{C}_{\kappa\kappa}^{ij}(\ell_1) ,\,\bm{C}_{\kappa\kappa}^{kl}(\ell_2))=&
    \frac{4\pi\delta_{\ell_1\ell_2}}{\Omega_\mathrm{s}(2\ell_1+1)\Delta\ell_1}[\bar{\bm{C}}_{\kappa\kappa}^{ik}(\ell_1) \bar{\bm{C}}_{\kappa\kappa}^{jl}(\ell_1)\\
    &+\bar{\bm{C}}_{\kappa\kappa}^{il}(\ell_1)\bar{\bm{C}}_{\kappa\kappa}^{jk}(\ell_1)],
\end{aligned}
\end{equation}
where
$
    \bar{\bm{C}}_{\kappa\kappa}^{ij}(\ell_1)\equiv
    \bm{C}_{\kappa\kappa}^{ij}(\ell_1)+\delta_{ij}\frac{\sigma^2_{\bm{\epsilon}}}{\bar{n}^{i}_\mrm{src}},
$ $\delta_{ij}$ is the Kronecker delta function, $\sigma_{\bm{\epsilon}}$ is the shape noise of the two shear components, $\bar{n}_{\mathrm{src}}^i$ is the mean source galaxy number density in the tomography bin $i$, $\Delta\ell_1$ is the $\ell_1$ bin width, $\Omega_\mathrm{s}$ is the survey area.
$\mat{C}_{\rm{NG}}$ is further decomposed into $\mat{C}_{\mrm{NG,\,0}}$ 
\begin{equation}
\label{eqn:C_NG0}
\begin{aligned}
    \mat{C}_{\mrm{NG,\,0}}&(\bm{C}_{\kappa\kappa}^{ij}(\ell_1), \bm{C}_{\kappa\kappa}^{kl}(\ell_2))=\frac{1}{\Omega_s}\int_{|\mathbf{l}|\in\ell_1}\frac{\dd^2 \mathbf{l}}{A(\ell_1)}\int_{|\mathbf{l}^\prime|\in\ell_2}\frac{\dd^2\mathbf{l}^\prime}{A(\ell_2)}\\
    &\times\int \dd\chi\frac{q_\kappa^{i}(\chi)q_\kappa^{j}(\chi)q_\kappa^{k}(\chi)q_\kappa^{l}(\chi)}{\chi^6}\\
    &\times T_\kappa^{ijkl}(\mathbf{l}/\chi,\, -\mathbf{l}/\chi,\, \mathbf{l}^\prime/\chi,\, -\mathbf{l}^\prime/\chi;\,z(\chi)),
\end{aligned}
\end{equation}
where $\int_{|\mathbf{l}|\in\ell_1}$ means the integration range is an annulus in Fourier space of radius $\ell_1$ and width $\Delta\ell_1$, $A(\ell_1)\equiv \int_{|\mathbf{l}|\in\ell_1} \mathrm{d}^2\mathbf{l}$ is the integration area in Fourier space, and $T_\kappa^{ijkl}$ is the trispectrum of cosmic shear~\citep{tj09}. Note that $\mat{C}_\mathrm{G}$ is free from the survey window effects. The super-sample covariance $\mat{C}_{\mrm{SSC}}$~\citep{TH13}
\begin{equation}
\label{eqn:C_SSC}
\begin{aligned}
    \mat{C}_{\mrm{SSC}}&(\bm{C}_{\kappa\kappa}^{ij}(\ell_1), \bm{C}_{\kappa\kappa}^{kl}(\ell_2))=\int \dd\chi\frac{q_\kappa^{i}(\chi)q_\kappa^{j}(\chi)q_\kappa^{k}(\chi)q_\kappa^{l}(\chi)}{\chi^4}\\
    &\times \frac{\partial P_{\kappa}(\ell_1/\chi,\,z(\chi))}{\partial \delta_b}\frac{\partial P_{\kappa}(\ell_2/\chi,\,z(\chi))}{\partial \delta_b}\sigma_b(\Omega_\mathrm{s};\,z(\chi)),
\end{aligned}
\end{equation}
which captures the uncertainties originating from the large-scale density modes outside the survey window. Here $\sigma_b(\Omega_\mathrm{s};\,z(\chi))$ is the variance of the background mode over the survey window, and $\frac{\partial P_{\kappa}(\ell_1/\chi,\,z(\chi))}{\partial \delta_b}$ captures the response of the 3D convergence power spectrum to the background density mode $\delta_b$. For more details, see the appendix in~\cite{Krause_Tim17} and reference therein.

\begin{figure*}
	\includegraphics[width=\columnwidth]{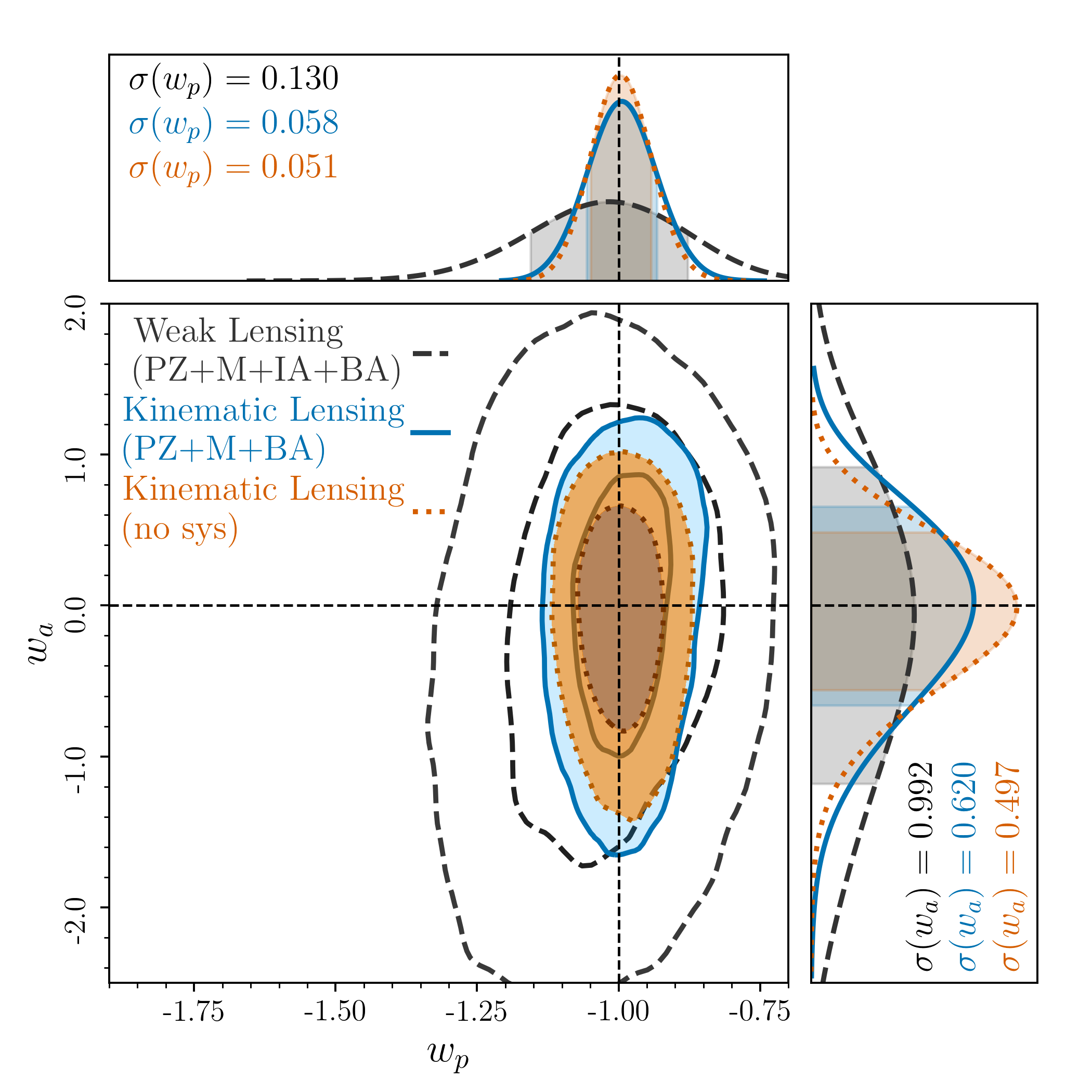}
	\includegraphics[width=\columnwidth]{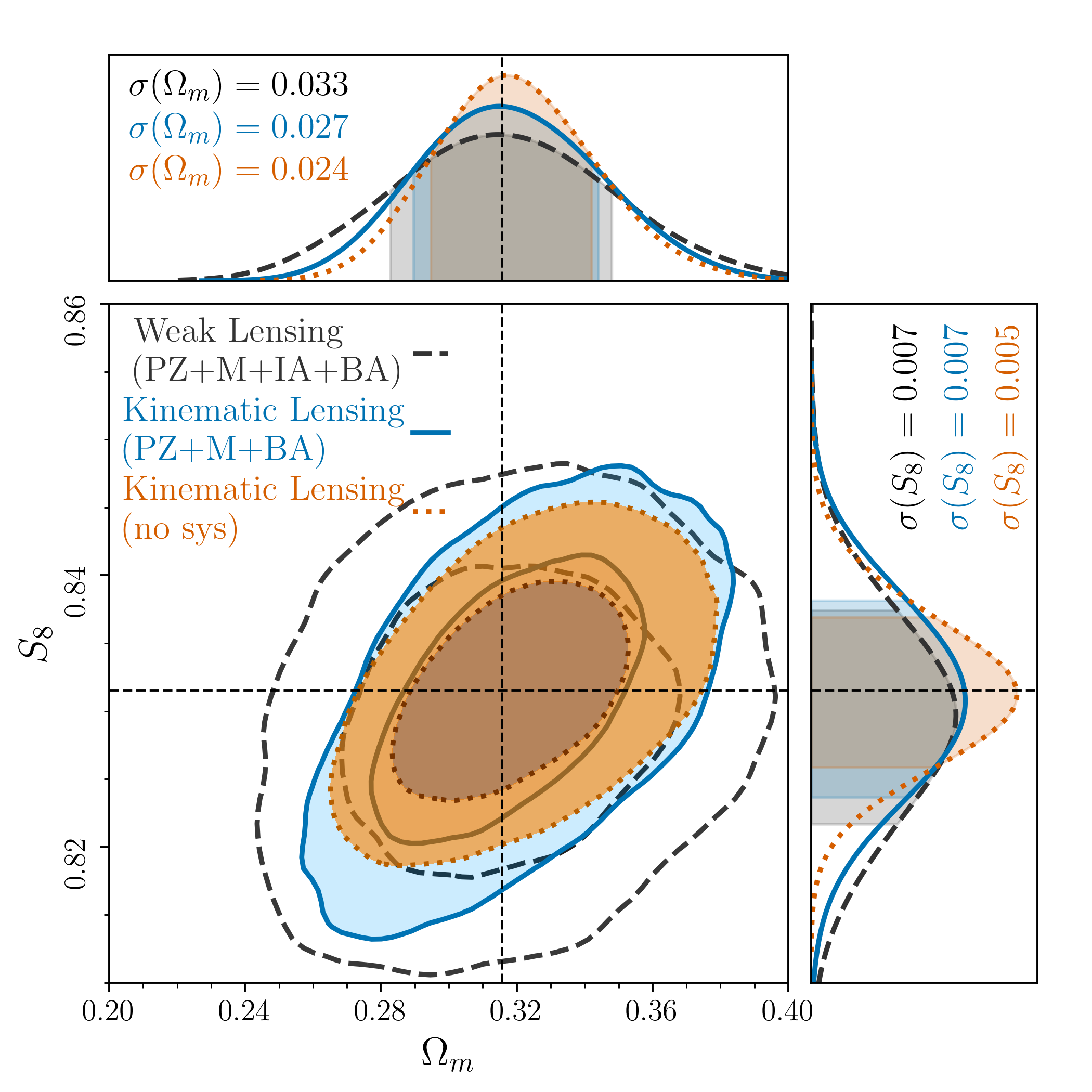}
    \caption{Posterior probability distribution of $w_p$-$w_a$(left) and $\Omega_\mrm{m}$-$S_8$(right). The black dashed lines are inferred from traditional weak lensing method. They are marginalized over PZ, M, IA and BA (see Sect. ~\ref{sec:sys} for the definitions of the abbreviations). The blue solid lines are kinematic lensing result marginalized over PZ, M, BA, but not IA. The orange dotted lines are inferred from kinematic lensing method, and assume fiducial nuisance parameters $\bm{p}_\mrm{nu}$, i.e. only sampling $\bm{p}_\mrm{co}$ without systematics modelling. The inner and outer contours show the $68$ per cent and $95$ per cent confidence level. Fiducial cosmological parameters are denoted as horizontal and vertical black dashed lines. In the 1D marginal probability distribution panels, the shaded area indicates the $68$ per cent confidence interval, with the 1$\sigma$ error bar annotated as figure legend, in the same colour scheme.}
    \label{fig:LCDM-WL-KL}
\end{figure*}

\subsection{Systematics Modelling}
\label{sec:sys}
We consider four types of systematic uncertainties in our modelling: redshift uncertainty, multiplicative shear calibration bias, intrinsic alignment and baryonic feedback. The related parameter settings are summarized in Table~\ref{tab:params_table}.
\begin{itemize}
    \item \textbf{Redshift Uncertainty (PZ)}: We assume that redshift errors follow Gaussian distributions for both the WL and the KL scenarios. A galaxy in sample $x$ and in the $i$-th redshift bin with redshift $z_{\mrm{true}}$ has a probability of $p^{i}(z_{\mrm{obs}}|\,z_{\mrm{true}},\,x)$ to be mis-identified to $z_{\mrm{obs}}$:
    \begin{equation}
    \begin{aligned}
        p^{i}(z_{\mrm{obs}}|\,z_{\mrm{true}},\,x)=&\frac{1}{\sqrt{2\pi}\sigma_{z,\,x}^{i}(1+z)}\times\\
        &\mrm{exp}\left[-\frac{(z_{\mrm{true}}-z_{\mrm{obs}}-\Delta_{z,\,x}^{i})^2}{2(\sigma_{z,\,x}^{i}(1+z))^2}\right],
    \end{aligned}
    \end{equation}
    $\Delta_{z,\,x}^{i}$ and $\sigma_{z,\,x}^{i}$ are the bias and the standard deviation of the observed redshift of galaxy sample $x$ in the $i$-th tomographic bin. Thus the photo-$z$ (or spec-$z$ for the KL sample) distribution for the $i$-th tomography bin is
    \begin{equation}
        n_x^{i}(z_\mrm{ph})=\int_{z_{\mrm{min},\,x}^{i}}^{z_{\mrm{max},\,x}^{i}} n_x(z)p^{i}(z_{\mrm{ph}}|\,z,\,x) \dd z.
    \end{equation}
    
    For the WL scenario, we assume the optimistic dataset from LSST Y10 result and \textit{Roman Space Telescope} HLIS full reference survey result, resulting to $\Delta_{z,\,\mrm{WL}}^{i}=0$ and $\sigma_{z,\,\mrm{WL}}^{i}=0.01$~\citep{WFIRST-LSST}. We also add our knowledge of photo-$z$ uncertainty in the likelihood simulation as a Gaussian prior,
    \begin{equation}
        \mrm{ln}(L_{\mrm{prior}}^{\mrm{photo}\textit{-}z})=-\frac{1}{2}\left[(\frac{\Delta_{z,\,x}^{i}-\bar{\Delta}_{z,\,x}^{i}}{\sigma(\Delta_{z,\,x}^{i})})^2+(\frac{\sigma_{z,\,x}^{i}-\bar{\sigma}_{z,\,x}^{i}}{\sigma(\sigma_{z,\,x}^{i})})^2\right],
    \end{equation}
    with $\Delta_{z,\,\mrm{WL}}^{i}\sim \mathcal{N}(0,\,0.001)$ and $\sigma_{z,\,\mrm{WL}}^{i}\sim\mathcal{N}(0.01,\,0.002)$.
    
    We motivate the redshift error of KL sample by the grism resolution
    %The redshift error of the KL sample is estimated from grism resolution 
    $\mathcal{R}^{-1}\approx0.002$. 
    We set $(\Delta_{z,\,\mrm{KL}}^{i},\,\sigma_{z,\,\mrm{KL}}^{i})=(0,\,0.002)$ and Gaussian priors $(\mathcal{N}(0,\,4\mrm{e}{-4}),\,\mathcal{N}(0.002,\,4\mrm{e}{-4}))$ for them correspondingly. 
    
    We note that while our WL redshift uncertainty is rather optimistic, our KL choice is very conservative, since with high SNR detection, the grism redshift uncertainty can be orders of magnitude smaller.
    
    \item \textbf{Multiplicative Shear Calibration Bias (M)}: Uncertainties in galaxy shape measurement come from a variety of sources: image noise, non-perfect galaxy flux profile model, non-linear effects in detector, atmospheric PSF, galaxy blending, etc. Those errors are summarized by the shear calibration bias parameters. We model the multiplicative shear calibration bias with $m^{i}$ (for the $i$-th redshift bin), which modifies the shear-shear angular power spectrum as
    \begin{equation}
        \bm{C}_{\kappa\kappa}^{ij}(\ell) \longrightarrow (1+m^{i})(1+m^{j})\bm{C}_{\kappa\kappa}^{ij}(\ell).
    \end{equation}
    
    We take the fiducial value $m^{i}=0$ and a Gaussian prior of $\mathcal{N}(0,\,0.002)$ for the WL sample, which is the same as \cite{WFIRST_MultiProbe,WFIRST-LSST}. Since the KL sample is much more brighter and has higher SNR than the WL sample, we set $m^{i}=0$ in the data vector and the prior of $m^{i}$ to be $\mathcal{N}(0,\,4\mathrm{e}{-4})$ for the KL sample. 
    We note that existing work on multiplicative shear calibration bias found that $m$ is a function of SNR and galaxy size, and galaxies of $\mathrm{SNR}\,\approx100$ have about 2 times smaller $m$ uncertainty than those of $\mathrm{SNR}\,\approx 10$~\citep{FHH+17,ZSS+18,MLL+18}. Our KL sample approximately follows a log-normal distribution in $r$-band SNR $\mathrm{log}_{10}(\mathrm{SNR})\approx\mathcal{N}(2.62, 0.29)$, while the above papers are focused on galaxies with $\mathrm{SNR}\,\in [10,200]$. This is significantly fainter than most objects in our galaxy sample, which motivates our choice of the KL sample to have 5 times lower $m$ uncertainty in this pilot forecast.
    
    \item \textbf{Intrinsic Alignment (IA)}: 
    
    The strength of intrinsic alignment of galaxies depends on the type of galaxy considered. A simple theoretical description is that elliptical galaxies are aligned by the tidal field caused by the host dark matter halo profile, while spiral galaxies are aligned by the tidal torquing generated by the host halo angular momentum, which is much weaker than the tidal alignment~\citep{Catelan01}.
    
    We model intrinsic alignment by including two extra shape correlation terms: The first term affects source galaxies at the same redshift bin $\bm{C}_{\mrm{II}}^{ii}(\ell)$, which models the shape correlation of background galaxies caused by their local tidal field, and the second affects galaxies in different redshift bins $\bm{C}_{\mrm{GI}}^{ij}(\ell)$, where the foreground large-scale structure correlates the orientation a foreground galaxy though tidal alignment or tidal torquing with a background galaxy's shear distortion. 
    \begin{equation}
        \bm{C}_{\kappa\kappa}^{ij}(\ell)\longrightarrow \bm{C}_{\kappa\kappa}^{ij}(\ell)+\bm{C}_{\mrm{II}}^{ij}(\ell)+\bm{C}_{\mrm{GI}}^{ij}(\ell).
    \end{equation}
    
    We adopt the Non-Linear Alignment description of IA ~\citep[NLA, see][]{his04,brk07,Krause16_IA} for red elliptical galaxies, and ignore the contribution from spiral galaxies~\citep[see][and reference therein for more details]{WFIRST-LSST}. We model uncertainties in the NLA model through four different parameters, which are $A_{\mrm{IA}}$ describing the amplitude of IA, $\beta_{\mrm{IA}}$ constraining the luminosity dependence, and $\eta_{\mrm{IA}},\,\eta_{\mrm{IA}}^{\mrm{high}\textit{-}z}$ depicting redshift dependence. The fiducial values and priors are shown in Table~\ref{tab:params_table}.
    
    However, as we have seen in equation~(\ref{eqn:reduced_shears}), the KL method is unaffected (to leading order) by the traditional IA contamination that plagues WL, since we can in principle isolate the intrinsic shape from the measured shape. Also, the KL sample is dominated by emission line galaxies which mostly have spiral morphology, thus the IA amplitude is expected to be order-of-magnitude smaller than that for the WL sample, which contains lots of elliptical galaxies. As a result, IA uncertainties are negligible for KL and we do not include IA in the KL scenarios. 
    
    \item \textbf{Baryonic Feedback (BA)}: As we approach small scales in cosmic shear, baryonic feedback effects during the formation and evolution of galaxies start to impact our observable through the reduced clustering of matter. 
    The resulting modification in the matter power spectrum is a mixture of two leading effects: the feedback distributes baryons to the outskirt of the halo thus suppresses the formation of structure, and the cooling of baryons causes more gravitational collapse towards the halo centre. 
    We adopt the principal component analysis technique~\citep[PCA,][]{hem19,ekd15} to mitigate the baryon feedback. The basic idea is to project the impact of baryon feedback on the matter power spectrum on to several principal components (PCs), utilizing a range of simulations that span a variety of baryon physics recipes. 
    We assume the real baryon feedback in the universe could be approximated by a linear combination of those PCs, and fit the amplitude $Q_{i}$ of $\mrm{PC}_{i}$. For a more detailed modelling description, see~\cite{WFIRST-LSST}.
    
    The simulations we select for the principal component space are Horizon-AGN~\citep{dpp16}, Illustris/IllustrisTNG~\citep{wsp18,psn18}, Eagle simulation~\citep{scb15}, Massiveblack-II~\citep{kdc15} and the OWLS AGN simulation~\citep{sdb10,dsb11}. 
    We include the first three PCs in the modelling and set the fiducial value of $Q_{i}$ to 0, given that we use a dark-matter-only $\bm{C}_{\kappa\kappa}^{ij}(\ell)$ as our fiducial input data vector. The priors for our baryon mitigation scheme are listed in Table~\ref{tab:params_table}.
\end{itemize}

%%% Results
%%% =======
\section{Simulated kinematic lensing analyses}
\label{sec:results}

In this section, we compare the constraining power of a variety of lensing scenarios, under different cosmology models. 
Our metrics of comparison are 2D posterior contours and the standard Figure-of-Merit (FoM, $\sqrt{\mrm{det}[\mat{C}_{p_1 p_2}^{-1}]}$ where $\mat{C}_{p_1 p_2}$ is the covariance matrix between parameter $p_1$ and $p_2$) of $\Omega_\mrm{m}$-$\sigma_8$ and/or $w_0$-$w_a$. 
To decorrelate parameters, we replace $\sigma_8$ with $S_8\equiv\sigma_8(\Omega_\mrm{m}/0.315)^{0.35}$ and $w_0$ with $w_p\equiv w_0+(1-1/(1+z_\mrm{p}))w_a$, where the pivot redshift $z_\mathrm{p}=0.4$ is chosen such that the uncertainty on $w(z_\mathrm{p})$ is minimal.

\subsection{Comparing Weak Lensing and Kinematic Lensing}
\label{sec:WL-KL}

We compare the constraining power of weak lensing and kinematic lensing method ($N_{\mrm{tomo}}=10$) using three different likelihood setups:
\begin{itemize}
    \item The WL scenario samples $\bm{p}_{\mrm{co}}$ and $\bm{p}_{\mrm{nu}}$, including PZ, M, IA and BA, sampling 35 parameters in total. 
    \item The KL scenario samples $\bm{p}_{\mrm{co}}$ and $\bm{p}_{\mrm{nu}}$ including PZ, M and BA, sampling 31 parameters in total.  
    \item The KL scenario samples $\bm{p}_{\mrm{co}}$ only, assuming fiducial $\bm{p}_{\mrm{nu}}$, which means it samples only 7 parameters only. The idea of this chain is to study the impact of nuisance parameters in the KL context.
\end{itemize}

We show the posteriors of $\Omega_\mrm{m}$-$S_8$ and $w_p$-$w_a$ inferred from those chains in Fig.~\ref{fig:LCDM-WL-KL}. Compared with traditional weak lensing (the black dashed contours), the kinematic lensing (the blue solid contours, with systematics mitigation) shows noticeable improvement in constraining power on $\Omega_\mrm{m}$-$S_8$ and significant enhancement on $w_p$-$w_a$. 
The improvements quantified by the FoM are shown in Table~\ref{tab:FoM}, and $\mrm{FoM}_{w_p\textit{-}w_a}$ has increased by 2.65 times by adopting KL method. 
Despite the decrease of effective galaxy number density, the excellent shape noise control benefits the dark energy science significantly.

\begin{figure}
	\includegraphics[width=\columnwidth]{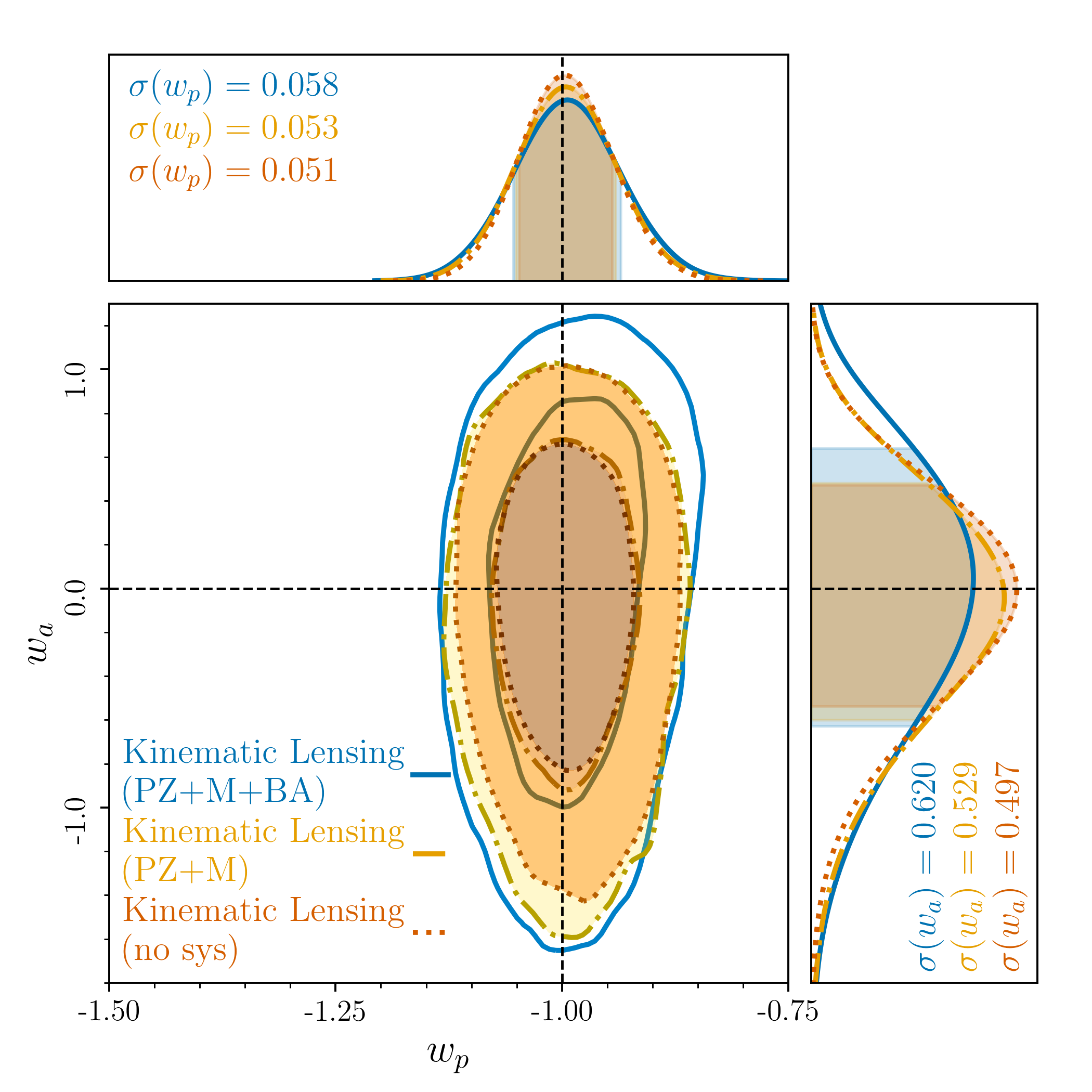}
    \caption{Comparison of the $w_p$-$w_a$ KL constraining power degradation from systematics. The systematics modelling includes: i) PZ, M and BA (the blue solid contour) ii) PZ and M (the yellow solid contour) and iii) no systematic modelling, assuming perfect knowledge of systematics (the orange dotted contour). All of the three chains assume KL with $N_{\mrm{tomo}}=10$ and $\sigma_{\bm{\epsilon}}=0.035$.}
    \label{fig:KL-sys-degrade}
\end{figure}

Another important conclusion is that the constraining power of KL is not systematics-limited given the current configuration of \rst\ HLS. From Table~\ref{tab:FoM}, marginalization over spec-$z$ uncertainty and shear calibration bias degrades the FoM by 19.7 per cent, while adding baryon feedback into marginalization brings another 15.7 per cent degradation (also see Fig.~\ref{fig:KL-sys-degrade}). We do not consider other systematics like galaxy blending in kinematic lensing but given the carefully selected high S/N sample, KL is likely more robust against imperfect systematics modelling compared to WL.  

\begin{table}
    \centering
    \begin{tabular}{l|c|ccc}
    \toprule 
     FoM & WL & \multicolumn{3}{c}{KL($N_{\mrm{tomo}}=10$)} \\
     \hline 
     \makecell{Systematics\\Mitigation}  & \makecell{PZ+M+\\IA+BA} & \makecell{PZ+M+\\BA} & PZ+M & no sys\\
     \midrule
    $\mrm{FoM}_{w_p\textit{-}w_a}$ & 10.55 & 38.51 & 47.85 & 59.58\\
    $\mrm{FoM}_{\Omega_\mrm{m}\textit{-}S_8}$ & 5307 & 9017 & 11533 & 13543\\
    \bottomrule
    \end{tabular}
    \caption{The FoMs of $w_p$-$w_a$ and $\Omega_\mrm{m}$-$S_8$, comparing between the WL method and KL method ($N_{\mrm{tomo}}=10$). We calculate the FoM by picking the top 68 per cent points with the highest likelihood in each of the chains, then estimate the covariance matrix from these sub-samples. The columns from left to right show the FoM of i) WL method marginalized over full systematics, including PZ, M, IA and BA (see Sect.~\ref{sec:sys} for the abbreviations and the modelling details) ii) KL method marginalized over PZ, M and BA iii) KL method marginalized over PZ and M, and iv) KL method sampling $\bm{p}_\mrm{co}$ only (no sys).}
    \label{tab:FoM}
\end{table}

\begin{figure*}
	\includegraphics[width=\columnwidth]{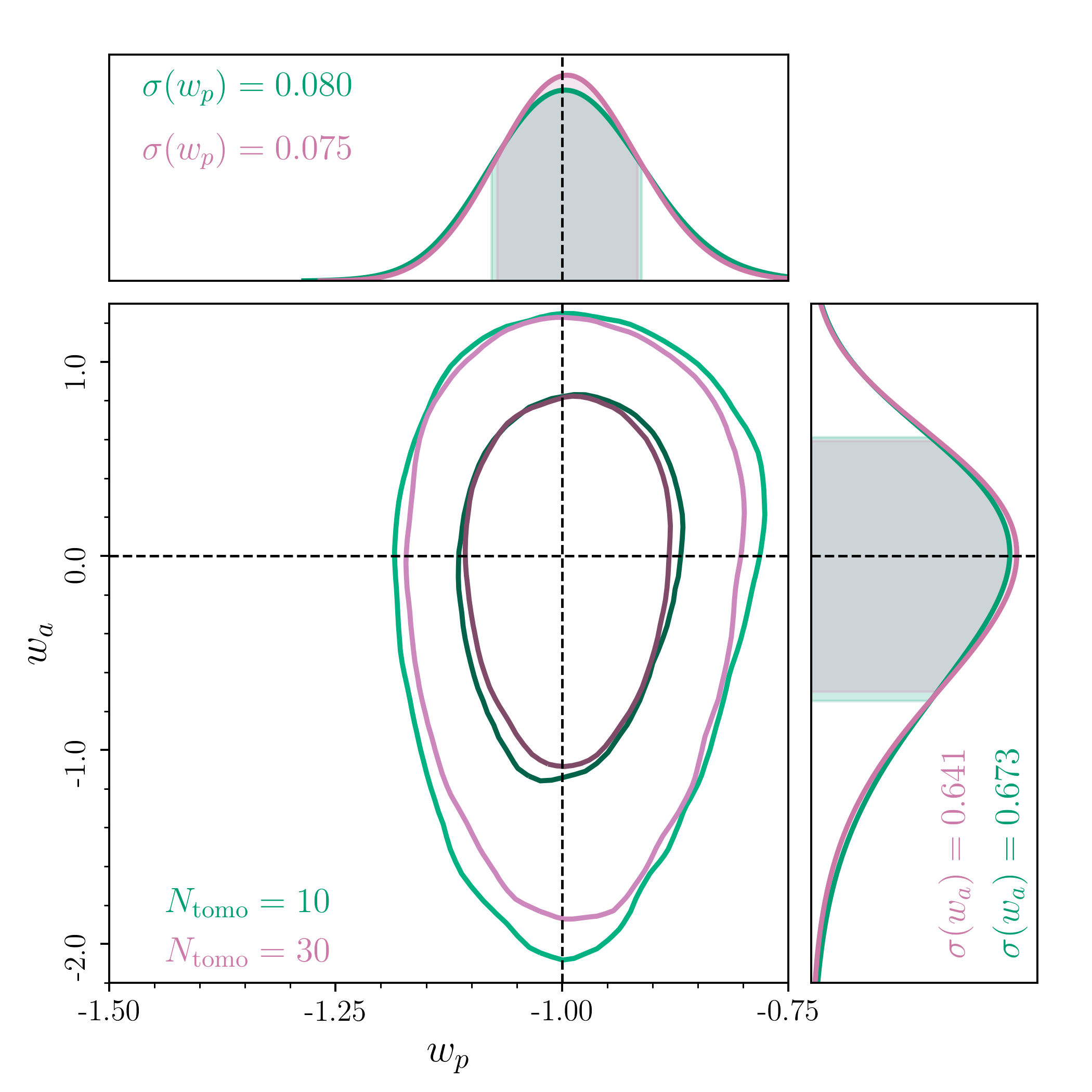}
	\includegraphics[width=\columnwidth]{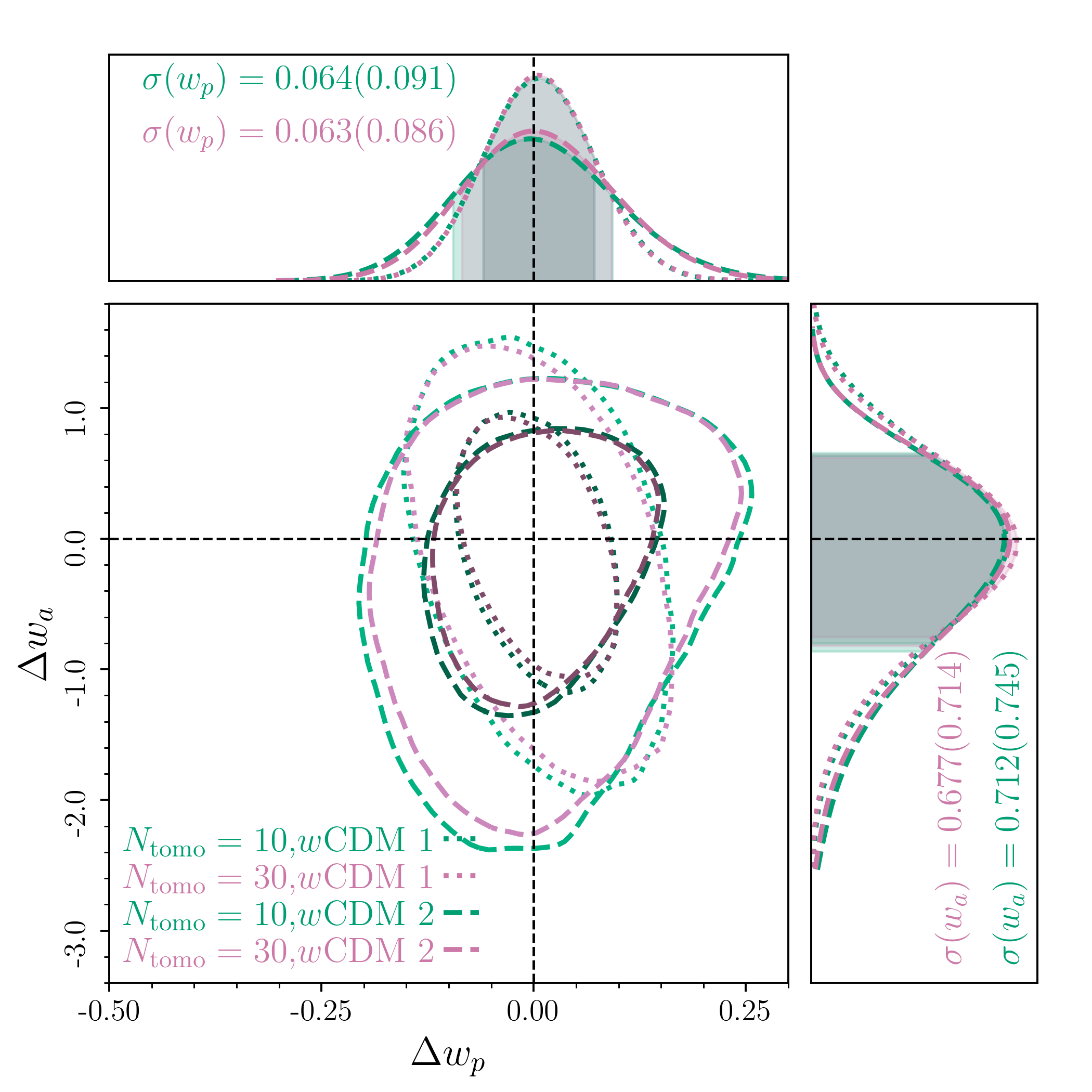}
    \caption{Comparing the constraining power on $w_p$-$w_a$ of kinematic lensing survey with different redshift binning strategies and cosmologies. 
    In the left panel we assume the fiducial cosmology to be $\Lambda$CDM, while in the right panel, we study the posterior distribution on differential dark energy EoS $\Delta w_p$-$\Delta w_a$ under two $w$CDM fiducial cosmologies, $w$CDM 1: $(w_p,w_a)=(-0.92,-2.21)$ or $(w_0,w_a)=(-0.289, -2.21)$ (the dotted contours, of which the 68 per cent C.L. are annotated without bracket in the 1D posteriors) and $w$CDM 2: $(w_p,w_a)=(-1.08,0.59)$ or $(w_0,w_a)=(-1.249, 0.59)$ (the dashed contours, of which the 68 per cent C.L. are annotated with brackets in the 1D posteriors). 
    The green and violet colours correspond to $N_{\mrm{tomo}}=10$ and $N_{\mrm{tomo}}=30$ scenarios. All the chains sample $\bm{p}_\mrm{co}$ only, and assume $\sigma_{\bm{\epsilon}}=0.056$.}
    \label{fig:KL-10-30}
\end{figure*}

\subsection{Kinematic Lensing 10 Tomography Bins v.s. 30 Tomography Bins}
\label{sec:KL10-30}

\begin{table}
    \centering
    \begin{tabular}{l|cc}
    \toprule 
      $N_{\mrm{tomo}}$ & 10 & 30\\
      %$\sigma_{\bm{\epsilon}}$ & 0.056 & 0.056\\
     \midrule
    $\mrm{FoM}_{w_p\textit{-} w_a}$ & 27.65 & 31.14\\% 59.58 & 27.65 & 31.14
    \bottomrule
    \end{tabular}
    \caption{The FoM of $w_p$-$w_a$, comparing between different tomography strategies of KL method. The columns from left to right show the FoM of $N_{\mrm{tomo}}=10$ and $N_{\mrm{tomo}}=30$. Both of them sample cosmological parameters only, and assume $\sigma_{\bm{\epsilon}}=0.056$.}
    \label{tab:FoM_10-30}
\end{table}

We compare the performance of 10 and 30 tomography bins KL methods assuming $\Lambda$CDM as the fiducial cosmology. 
We run two chains with one assuming 10 tomography bins and the other one assuming 30.
As we mentioned in Sect.~\ref{sec:shape_noise}, we increase the shape noise of the $N_{\mrm{tomo}}=30$ chain to 0.056 to make the covariance matrix positive-definite. We adopt the same setting for the $N_{\mrm{tomo}}=10$ chain for a fair comparison between the two tomography schemes.
Both chains are sampling $\bm{p}_\mrm{co}$ only. For each chain, 560 walkers are deployed and each walker makes 5,000 steps. 
We drop the first 0.5M steps as burn-in for each of those. 
The posterior distributions on $w_p$-$w_a$ are shown in the left panel of Fig.~\ref{fig:KL-10-30}, and the FoM of $w_p$-$w_a$ are shown in Table~\ref{tab:FoM_10-30}. 

The $N_{\mrm{tomo}}=30$ scenario is marginally better than the $N_{\mrm{tomo}}=10$ one (12.6 per cent increase). 
There could be two possible reasons for the in-sensitiveness. 
The first one is that the observables $C_{\kappa\kappa}^{ij}(\ell)$ are featureless and smooth functions of redshift and scale, hence 10 tomography bins are sufficient to capture most of the information encoded in shear-shear angular power spectrum.
Another possible reason is that in the fiducial $\Lambda$CDM cosmology, dark energy is constant over all the redshift bins, and there is no room for more flexible tomography evolution given the allowed range of $w_p$-$w_a$. We conclude that 10 tomography bins are sufficient to capture the cosmological information in the case considered, but the same must not be true when stronger dark energy evolution across tomography bins is allowed. 

In order to discriminate between the two scenarios we explore fiducial data vectors computed from a cosmology with time-evolving dark energy , i.e. $w_a\neq0$. According to the \textit{Planck} 2018 result~\citep[fig.~30 in ][]{Planck2018}, we choose two sets of fiducial parameters: $(w_0,\,w_a)=(-0.289,\,-2.21)$ ($w$CDM 1) and $(w_0,\,w_a)=(-1.249,\,0.59)$ ($w$CDM 2), which correspond to two opposite corners at the edge of $2\sigma$ contour inferred from TT,TE,EE+lowE+lensing+BAO/RSD+WL. 
For each set of parameters, we run two chains as we do in the $\Lambda$CDM cases. We show the results in the right panel in Fig.~\ref{fig:KL-10-30}, with the dotted contours assuming fiducial $(w_0,\,w_a)=(-0.289,\,-2.21)$ and the dashed contours assuming fiducial $(w_0,\,w_a)=(-1.249,\,0.59)$. From the posteriors we only see tiny differences between the 10 bins cases and the 30 bins cases (note that these assume the same shape noise level). Thus for scenarios with evolving dark energy within the 1$\sigma$ region allowed by the \textit{Planck} CMB experiment, we can not gain further constraining power by increasing the number of tomographic bins. The observables $\bm{C}_{\kappa\kappa}^{ij}(\ell)$ are smooth in nature and 10 tomography bins are enough to capture most of the information. We note however that more complex dark energy scenarios might benefit from this narrow tomography idea that is feasible with KL.

\section{Conclusions}
\label{sec:conclusions}
%The last numbered section should briefly summarise what has been done, and describe the final conclusions which the authors draw from their work.

% Roman summary
In this work we study the performance of a kinematic lensing survey conducted with the \rst. 
In its reference design the High Latitude Survey covers 2,000 square degrees in the 5-year prime mission with imaging and spectroscopy. 
The HLIS will map objects in $Y$, $J$, $H$ and $F184$ bands to a depth of $\mathrm{M}_{\mrm{AB,}J}=26.7$, and the HLSS is targeting emission line galaxies at redshift 0.524--2.855, with 7$\sigma$ flux limit of $1.0\times 10^{-16}\,(1.8\times 10^{-16})\,\mrm{erg}\,\mrm{cm}^{-2}\,\mrm{s}^{-1}$ for point source (extended source), with grism resolution $\mathcal{R}=461(\frac{\lambda}{1\,\mrm{\mu m}})$. 
% source selection: number
Based on the CANDELS and CMC catalogue we have defined realistic galaxy samples and we get a source galaxy number density of $51\ \mrm{gal\,arcmin}^{-2}$ for weak lensing and $4\ \mrm{gal\,arcmin}^{-2}$ for kinematic lensing (assuming 50 per cent grism success rate). We note that the KL sample number density estimation is also supported by literature using multiple independent methods, like EL-COSMO mock catalogue, SAM-based mock catalogue and luminosity functions.
To estimate the effective shape noise of the KL sample, we design a toy-model for the KL shear estimation process with the \rst\ and solve for the MLE shear estimator drawing from the said model. The resulting shape noise ($\sigma_{\bm{\epsilon}}^\mrm{KL}=0.035$) shows a significant improvement over the one in traditional weak lensing ($\sigma_{\bm{\epsilon}}^\mrm{WL}=0.37$).

We demonstrate KL's constraining power by simulating shear-shear only likelihood analyses and comparing to the traditional WL case. For WL, we model a full suite of systematic uncertainties such as photo-$z$ uncertainty, multiplicative shear calibration bias, intrinsic alignment and baryonic feedback. For KL, IA mitigation is unnecessary since the contamination from IA can be calibrated and removed during the isolation the intrinsic shape information in the KL measurement process. 

Other prominent WL systematics also experience substantial improvement in the KL case, e.g. redshift uncertainties are insignificant given that we have spectroscopic information, and multiplicative shear calibration bias uncertainties are greatly diminished given the high S/N imaging sample.

Comparing to the standard WL strategy, we find significant improvement in constraining power on $w_p$-$w_a$ (FoM$_{\mrm{KL}}$=3.65 FoM$_{\mrm{WL}}$) and noticeable enhancement on $\Omega_\mrm{m}$-$S_8$ (FoM$_{\mrm{KL}}$=1.70 FoM$_{\mrm{WL}}$). 
We also quantify how the FoM$_{\mrm{KL}}$ degrades with different systematics. Compared with the fiducial KL case where no systematics are included, baryonic feedback causes a 15.68 per cent degradation. Redshift uncertainty and shear calibration bias in total contribute another 19.69 per cent degradation in the $w_p$-$w_a$ FoM. Hence, with the current instrument capability of \rst, the shear calibration bias and redshift uncertainty are comparable with the baryon feedback.

We also explore how increasing the number of tomographic redshift bins, which is an interesting concept given the reduced shape noise, affects the constraining power on $w_p$-$w_a$. Specifically, we increase the number of tomographic bins from our default of 10 to 30, which significantly increases the size of the covariance matrix. As fiducial input scenarios we consider three dark energy EoS scenarios ($w_0$,$\,w_a$)=(0.0,$\,$0.0)/(-0.289,$\,$-2.21)/(-1.249,$\,$0.59), that are consistent (but at the limits) of the \textit{Planck} 2018 TT,TE,EE+lowE+lensing+BAO/RSD+WL result. We find that no matter whether $w_{\mrm{DE}}$ is evolving or not, increasing the number of tomographic bins does not bring extra information about the EoS of dark energy. 
This can be explained by the smoothness nature of $\bm{C}_{\kappa\kappa}^{ij}(\ell)$ with regard to scale and redshift when assuming the standard $w_p$-$w_a$ dark energy parameterization. We note however that more complex dark energy or modified gravity scenarios will likely benefit from the more fine binning in redshift space enabled by KL. 

Generally speaking the decrease in shape noise using the KL method changes the intuition of how to optimize cosmological information as a function of survey and systematics mitigation strategy. As shown, intrinsic alignment, photo-$z$ and shear calibration systematics are less worrisome for KL surveys. In contrast, accessing and modelling small scale information, which are already important for WL, will likely become more impactful in the KL case. Also, compared to WL, increasing the survey area will be more important for KL given that the dominant statistical uncertainty for KL is cosmic variance. In particular for \rst, where the current reference survey envisions 2,000 deg$^2$, it will be important to explore a wide area KL survey \citep[similar to][]{WFIRST-LSST}, which would also benefit from overlap with ground-based spectroscopic resources.  

In summary, we consider KL with the \rst\ to be an extremely promising science avenue. It has different challenges than WL (rotation velocity modelling, Tully Fisher Relation uncertainties), but these are likely outweighed by the increase in statistical power and systematics robustness. In this context it is important to note that KL and WL data vectors can be combined in a likelihood analysis; the question is not to replace one with the other but to optimize the joint analysis. Similarly, the combination of KL with other probes such as galaxy clustering, peculiar velocities, CMB lensing, and SZ measurements are exciting opportunities. Lastly, we would like to point out that exploring KL in the context of galaxy cluster mass measurements is an important concept to explore in general and for \rst~in particular.

\section*{Acknowledgements}
%\updates{J. X. would gratefully acknowledge the helpful comments from Yun Wang, Christopher Hirata, David Weinberg, and Olivier Dor\'e.} 
This work is supported by NASA ROSES ATP 17-ATP17-0173 and NASA 15-WFIRST15-0008 grants. Simulations in this paper use High Performance Computing (HPC) resources supported by the University of Arizona TRIF, UITS, and RDI and maintained by the UA Research Technologies department.

\section*{Data Availability}
The data underlying this article will be shared on reasonable request to the corresponding author.
%%%%%%%%%%%%%%%%%%%%%%%%%%%%%%%%%%%%%%%%%%%%%%%%%%



\bibliographystyle{mnras}
\bibliography{example} % if your bibtex file is called example.bib

\begin{thebibliography}{}
\makeatletter
\relax
\def\mn@urlcharsother{\let\do\@makeother \do\$\do\&\do\#\do\^\do\_\do\%\do\~}
\def\mn@doi{\begingroup\mn@urlcharsother \@ifnextchar [ {\mn@doi@}
  {\mn@doi@[]}}
\def\mn@doi@[#1]#2{\def\@tempa{#1}\ifx\@tempa\@empty \href
  {http://dx.doi.org/#2} {doi:#2}\else \href {http://dx.doi.org/#2} {#1}\fi
  \endgroup}
\def\mn@eprint#1#2{\mn@eprint@#1:#2::\@nil}
\def\mn@eprint@arXiv#1{\href {http://arxiv.org/abs/#1} {{\tt arXiv:#1}}}
\def\mn@eprint@dblp#1{\href {http://dblp.uni-trier.de/rec/bibtex/#1.xml}
  {dblp:#1}}
\def\mn@eprint@#1:#2:#3:#4\@nil{\def\@tempa {#1}\def\@tempb {#2}\def\@tempc
  {#3}\ifx \@tempc \@empty \let \@tempc \@tempb \let \@tempb \@tempa \fi \ifx
  \@tempb \@empty \def\@tempb {arXiv}\fi \@ifundefined
  {mn@eprint@\@tempb}{\@tempb:\@tempc}{\expandafter \expandafter \csname
  mn@eprint@\@tempb\endcsname \expandafter{\@tempc}}}

\bibitem[\protect\citeauthoryear{{Abbott} et~al.,}{{Abbott}
  et~al.}{2018}]{DES_Y1_3x2pt}
{Abbott} T.~M.~C.,  et~al., 2018, \mn@doi [\prd] {10.1103/PhysRevD.98.043526},
  \href {https://ui.adsabs.harvard.edu/abs/2018PhRvD..98d3526A} {98, 043526}

\bibitem[\protect\citeauthoryear{{Abbott} et~al.,}{{Abbott}
  et~al.}{2019}]{DESY1xPlanckxSPT_6x2pt}
{Abbott} T.~M.~C.,  et~al., 2019, \mn@doi [\prd] {10.1103/PhysRevD.100.023541},
  \href {https://ui.adsabs.harvard.edu/abs/2019PhRvD.100b3541A} {100, 023541}

\bibitem[\protect\citeauthoryear{{Akeson} et~al.,}{{Akeson}
  et~al.}{2019}]{Roman_general}
{Akeson} R.,  et~al., 2019, arXiv e-prints, \href
  {https://ui.adsabs.harvard.edu/abs/2019arXiv190205569A} {p. arXiv:1902.05569}

\bibitem[\protect\citeauthoryear{{Amon} et~al.,}{{Amon}
  et~al.}{2021}]{DES_Y3_WL1}
{Amon} A.,  et~al., 2021, arXiv e-prints, \href
  {https://ui.adsabs.harvard.edu/abs/2021arXiv210513543A} {p. arXiv:2105.13543}

\bibitem[\protect\citeauthoryear{{Asgari} et~al.,}{{Asgari}
  et~al.}{2021}]{KiDS_shear20}
{Asgari} M.,  et~al., 2021, \mn@doi [\aap] {10.1051/0004-6361/202039070}, \href
  {https://ui.adsabs.harvard.edu/abs/2021A&A...645A.104A} {645, A104}

\bibitem[\protect\citeauthoryear{{Bartelmann} \& {Schneider}}{{Bartelmann} \&
  {Schneider}}{2001}]{BS01_review}
{Bartelmann} M.,  {Schneider} P.,  2001, \mn@doi [\physrep]
  {10.1016/S0370-1573(00)00082-X}, \href
  {https://ui.adsabs.harvard.edu/abs/2001PhR...340..291B} {340, 291}

\bibitem[\protect\citeauthoryear{{Beck}}{{Beck}}{2007}]{Beck07}
{Beck} R.,  2007, \mn@doi [\aap] {10.1051/0004-6361:20066988}, \href
  {https://ui.adsabs.harvard.edu/abs/2007A&A...470..539B} {470, 539}

\bibitem[\protect\citeauthoryear{{Bernstein} \& {Jarvis}}{{Bernstein} \&
  {Jarvis}}{2002}]{BJ02}
{Bernstein} G.~M.,  {Jarvis} M.,  2002, \mn@doi [\aj] {10.1086/338085}, \href
  {https://ui.adsabs.harvard.edu/abs/2002AJ....123..583B} {123, 583}

\bibitem[\protect\citeauthoryear{{Blain}}{{Blain}}{2002}]{Blain02}
{Blain} A.~W.,  2002, \mn@doi [\apjl] {10.1086/341103}, \href
  {https://ui.adsabs.harvard.edu/abs/2002ApJ...570L..51B} {570, L51}

\bibitem[\protect\citeauthoryear{{Bridle} \& {King}}{{Bridle} \&
  {King}}{2007}]{brk07}
{Bridle} S.,  {King} L.,  2007, \mn@doi [New Journal of Physics]
  {10.1088/1367-2630/9/12/444}, \href
  {http://adsabs.harvard.edu/abs/2007NJPh....9..444B} {9, 444}

\bibitem[\protect\citeauthoryear{{Brown} \& {Battye}}{{Brown} \&
  {Battye}}{2011a}]{BB_11a}
{Brown} M.~L.,  {Battye} R.~A.,  2011a, \mn@doi [\mnras]
  {10.1111/j.1365-2966.2010.17583.x}, \href
  {https://ui.adsabs.harvard.edu/abs/2011MNRAS.410.2057B} {410, 2057}

\bibitem[\protect\citeauthoryear{{Brown} \& {Battye}}{{Brown} \&
  {Battye}}{2011b}]{BB_11b}
{Brown} M.~L.,  {Battye} R.~A.,  2011b, \mn@doi [\apjl]
  {10.1088/2041-8205/735/1/L23}, \href
  {https://ui.adsabs.harvard.edu/abs/2011ApJ...735L..23B} {735, L23}

\bibitem[\protect\citeauthoryear{{Bull}, {Harrison}  \& {Huff}}{{Bull}
  et~al.}{2018}]{BHH_18}
{Bull} P.,  {Harrison} I.,   {Huff} E.,  2018, in {Murphy} E.,  ed.,
  Astronomical Society of the Pacific Conference Series Vol. 517, Science with
  a Next Generation Very Large Array. p.~803 (\mn@eprint {arXiv} {1806.08339})

\bibitem[\protect\citeauthoryear{{Camera}, {Harrison}, {Bonaldi}  \&
  {Brown}}{{Camera} et~al.}{2017}]{SKA_WL_3}
{Camera} S.,  {Harrison} I.,  {Bonaldi} A.,   {Brown} M.~L.,  2017, \mn@doi
  [\mnras] {10.1093/mnras/stw2688}, \href
  {https://ui.adsabs.harvard.edu/abs/2017MNRAS.464.4747C} {464, 4747}

\bibitem[\protect\citeauthoryear{{Catelan}, {Kamionkowski}  \&
  {Blandford}}{{Catelan} et~al.}{2001}]{Catelan01}
{Catelan} P.,  {Kamionkowski} M.,   {Blandford} R.~D.,  2001, \mn@doi [\mnras]
  {10.1046/j.1365-8711.2001.04105.x}, \href
  {https://ui.adsabs.harvard.edu/abs/2001MNRAS.320L...7C} {320, L7}

\bibitem[\protect\citeauthoryear{{Chang} et~al.,}{{Chang}
  et~al.}{2013}]{Chang13}
{Chang} C.,  et~al., 2013, \mn@doi [\mnras] {10.1093/mnras/stt1156}, \href
  {https://ui.adsabs.harvard.edu/abs/2013MNRAS.434.2121C} {434, 2121}

\bibitem[\protect\citeauthoryear{{DiGiorgio}, {Bundy}, {Westfall}, {Leauthaud}
  \& {Stark}}{{DiGiorgio} et~al.}{2021}]{DiGiorgio21}
{DiGiorgio} B.,  {Bundy} K.,  {Westfall} K.~B.,  {Leauthaud} A.,   {Stark} D.,
  2021, \mn@doi [\apj] {10.3847/1538-4357/ac2572}, \href
  {https://ui.adsabs.harvard.edu/abs/2021ApJ...922..116D} {922, 116}

\bibitem[\protect\citeauthoryear{{Ding}, {Seo}, {Huff}, {Saito}  \&
  {Clowe}}{{Ding} et~al.}{2019}]{DSH+19}
{Ding} Z.,  {Seo} H.-J.,  {Huff} E.,  {Saito} S.,   {Clowe} D.,  2019, \mn@doi
  [\mnras] {10.1093/mnras/stz1257}, \href
  {https://ui.adsabs.harvard.edu/abs/2019MNRAS.487..253D} {487, 253}

\bibitem[\protect\citeauthoryear{{Dubois}, {Peirani}, {Pichon}, {Devriendt},
  {Gavazzi}, {Welker}  \& {Volonteri}}{{Dubois} et~al.}{2016}]{dpp16}
{Dubois} Y.,  {Peirani} S.,  {Pichon} C.,  {Devriendt} J.,  {Gavazzi} R.,
  {Welker} C.,   {Volonteri} M.,  2016, \mn@doi [\mnras]
  {10.1093/mnras/stw2265}, \href
  {https://ui.adsabs.harvard.edu/abs/2016MNRAS.463.3948D} {463, 3948}

\bibitem[\protect\citeauthoryear{{Eifler}, {Schneider}  \& {Hartlap}}{{Eifler}
  et~al.}{2009}]{ESH_09}
{Eifler} T.,  {Schneider} P.,   {Hartlap} J.,  2009, \mn@doi [\aap]
  {10.1051/0004-6361/200811276}, \href
  {https://ui.adsabs.harvard.edu/abs/2009A&A...502..721E} {502, 721}

\bibitem[\protect\citeauthoryear{{Eifler}, {Krause}, {Dodelson}, {Zentner},
  {Hearin}  \& {Gnedin}}{{Eifler} et~al.}{2015}]{ekd15}
{Eifler} T.,  {Krause} E.,  {Dodelson} S.,  {Zentner} A.~R.,  {Hearin} A.~P.,
  {Gnedin} N.~Y.,  2015, \mn@doi [\mnras] {10.1093/mnras/stv2000}, \href
  {https://ui.adsabs.harvard.edu/abs/2015MNRAS.454.2451E} {454, 2451}

\bibitem[\protect\citeauthoryear{{Eifler} et~al.,}{{Eifler}
  et~al.}{2021a}]{WFIRST-LSST}
{Eifler} T.,  et~al., 2021a, \mn@doi [\mnras] {10.1093/mnras/stab533}, \href
  {https://ui.adsabs.harvard.edu/abs/2021MNRAS.507.1514E} {507, 1514}

\bibitem[\protect\citeauthoryear{{Eifler} et~al.,}{{Eifler}
  et~al.}{2021b}]{WFIRST_MultiProbe}
{Eifler} T.,  et~al., 2021b, \mn@doi [\mnras] {10.1093/mnras/stab1762}, \href
  {https://ui.adsabs.harvard.edu/abs/2021MNRAS.507.1746E} {507, 1746}

\bibitem[\protect\citeauthoryear{{Fenech Conti}, {Herbonnet}, {Hoekstra},
  {Merten}, {Miller}  \& {Viola}}{{Fenech Conti} et~al.}{2017}]{FHH+17}
{Fenech Conti} I.,  {Herbonnet} R.,  {Hoekstra} H.,  {Merten} J.,  {Miller} L.,
    {Viola} M.,  2017, \mn@doi [\mnras] {10.1093/mnras/stx200}, \href
  {https://ui.adsabs.harvard.edu/abs/2017MNRAS.467.1627F} {467, 1627}

\bibitem[\protect\citeauthoryear{{Fonseca} \& {Camera}}{{Fonseca} \&
  {Camera}}{2020}]{FC20}
{Fonseca} J.,  {Camera} S.,  2020, \mn@doi [\mnras] {10.1093/mnras/staa1136},
  \href {https://ui.adsabs.harvard.edu/abs/2020MNRAS.495.1340F} {495, 1340}

\bibitem[\protect\citeauthoryear{{Foreman-Mackey}, {Hogg}, {Lang}  \&
  {Goodman}}{{Foreman-Mackey} et~al.}{2013}]{emcee}
{Foreman-Mackey} D.,  {Hogg} D.~W.,  {Lang} D.,   {Goodman} J.,  2013, \mn@doi
  [\pasp] {10.1086/670067}, \href
  {https://ui.adsabs.harvard.edu/abs/2013PASP..125..306F} {125, 306}

\bibitem[\protect\citeauthoryear{Goodman \& Weare}{Goodman \&
  Weare}{2010}]{goodman2010}
Goodman J.,  Weare J.,  2010, \mn@doi [Commun. Appl. Math. Comput. Sci.]
  {10.2140/camcos.2010.5.65}, 5, 65

\bibitem[\protect\citeauthoryear{{Grogin} et~al.,}{{Grogin}
  et~al.}{2011}]{CANDELS}
{Grogin} N.~A.,  et~al., 2011, \mn@doi [\apjs] {10.1088/0067-0049/197/2/35},
  \href {https://ui.adsabs.harvard.edu/abs/2011ApJS..197...35G} {197, 35}

\bibitem[\protect\citeauthoryear{{Gurri}, {Taylor}  \& {Fluke}}{{Gurri}
  et~al.}{2020}]{GTF20}
{Gurri} P.,  {Taylor} E.~N.,   {Fluke} C.~J.,  2020, \mn@doi [\mnras]
  {10.1093/mnras/staa2893}, \href
  {https://ui.adsabs.harvard.edu/abs/2020MNRAS.499.4591G} {499, 4591}

\bibitem[\protect\citeauthoryear{{Gurri}, {Taylor}  \& {Fluke}}{{Gurri}
  et~al.}{2021}]{GTF21}
{Gurri} P.,  {Taylor} E.~N.,   {Fluke} C.~J.,  2021, \mn@doi [\mnras]
  {10.1093/mnras/stab423}, \href
  {https://ui.adsabs.harvard.edu/abs/2021MNRAS.502.5612G} {502, 5612}

\bibitem[\protect\citeauthoryear{{Hamana} et~al.,}{{Hamana}
  et~al.}{2020}]{HSC_Y1_2PCF}
{Hamana} T.,  et~al., 2020, \mn@doi [\pasj] {10.1093/pasj/psz138}, \href
  {https://ui.adsabs.harvard.edu/abs/2020PASJ...72...16H} {72, 16}

\bibitem[\protect\citeauthoryear{{Han}, {Manchester}  \& {Qiao}}{{Han}
  et~al.}{1999}]{HMQ_99}
{Han} J.~L.,  {Manchester} R.~N.,   {Qiao} G.~J.,  1999, \mn@doi [\mnras]
  {10.1046/j.1365-8711.1999.02544.x}, \href
  {https://ui.adsabs.harvard.edu/abs/1999MNRAS.306..371H} {306, 371}

\bibitem[\protect\citeauthoryear{{Hemmati} et~al.,}{{Hemmati}
  et~al.}{2019}]{hemmati2019}
{Hemmati} S.,  et~al., 2019, \mn@doi [\apj] {10.3847/1538-4357/ab1be5}, \href
  {https://ui.adsabs.harvard.edu/abs/2019ApJ...877..117H} {877, 117}

\bibitem[\protect\citeauthoryear{{Heymans} et~al.,}{{Heymans}
  et~al.}{2021}]{KiDS1000_3x2pt}
{Heymans} C.,  et~al., 2021, \mn@doi [\aap] {10.1051/0004-6361/202039063},
  \href {https://ui.adsabs.harvard.edu/abs/2021A&A...646A.140H} {646, A140}

\bibitem[\protect\citeauthoryear{{Hikage} et~al.,}{{Hikage}
  et~al.}{2019}]{SubaruHSC_shear19}
{Hikage} C.,  et~al., 2019, \mn@doi [\pasj] {10.1093/pasj/psz010}, \href
  {https://ui.adsabs.harvard.edu/abs/2019PASJ...71...43H} {71, 43}

\bibitem[\protect\citeauthoryear{{Hirata}}{{Hirata}}{2009}]{H09}
{Hirata} C.~M.,  2009, \mn@doi [\mnras] {10.1111/j.1365-2966.2009.15353.x},
  \href {https://ui.adsabs.harvard.edu/abs/2009MNRAS.399.1074H} {399, 1074}

\bibitem[\protect\citeauthoryear{{Hirata} \& {Seljak}}{{Hirata} \&
  {Seljak}}{2004}]{his04}
{Hirata} C.~M.,  {Seljak} U.,  2004, \mn@doi [\prd]
  {10.1103/PhysRevD.70.063526}, \href
  {http://adsabs.harvard.edu/abs/2004PhRvD..70f3526H} {70, 063526}

\bibitem[\protect\citeauthoryear{{Hirata}, {Gehrels}, {Kneib}, {Kruk},
  {Rhodes}, {Wang}  \& {Zoubian}}{{Hirata} et~al.}{2012}]{ETC}
{Hirata} C.~M.,  {Gehrels} N.,  {Kneib} J.-P.,  {Kruk} J.,  {Rhodes} J.,
  {Wang} Y.,   {Zoubian} J.,  2012, arXiv e-prints, \href
  {https://ui.adsabs.harvard.edu/abs/2012arXiv1204.5151H} {p. arXiv:1204.5151}

\bibitem[\protect\citeauthoryear{{Hu} \& {Jain}}{{Hu} \& {Jain}}{2004}]{HB04}
{Hu} W.,  {Jain} B.,  2004, \mn@doi [\prd] {10.1103/PhysRevD.70.043009}, \href
  {https://ui.adsabs.harvard.edu/abs/2004PhRvD..70d3009H} {70, 043009}

\bibitem[\protect\citeauthoryear{{Huang}, {Eifler}, {Mandelbaum}  \&
  {Dodelson}}{{Huang} et~al.}{2019}]{hem19}
{Huang} H.-J.,  {Eifler} T.,  {Mandelbaum} R.,   {Dodelson} S.,  2019, \mn@doi
  [\mnras] {10.1093/mnras/stz1714}, \href
  {https://ui.adsabs.harvard.edu/abs/2019MNRAS.488.1652H} {488, 1652}

\bibitem[\protect\citeauthoryear{{Huff}, {Krause}, {Eifler}, {Fang}, {George}
  \& {Schlegel}}{{Huff} et~al.}{2013}]{Eric13}
{Huff} E.~M.,  {Krause} E.,  {Eifler} T.,  {Fang} X.,  {George} M.~R.,
  {Schlegel} D.,  2013, arXiv e-prints, \href
  {https://ui.adsabs.harvard.edu/abs/2013arXiv1311.1489H} {p. arXiv:1311.1489}

\bibitem[\protect\citeauthoryear{{Joudaki} et~al.,}{{Joudaki}
  et~al.}{2020}]{KiDS+VIKING450xDES-Y1_shear}
{Joudaki} S.,  et~al., 2020, \mn@doi [\aap] {10.1051/0004-6361/201936154},
  \href {https://ui.adsabs.harvard.edu/abs/2020A&A...638L...1J} {638, L1}

\bibitem[\protect\citeauthoryear{{Jouvel} et~al.,}{{Jouvel} et~al.}{2009}]{CMC}
{Jouvel} S.,  et~al., 2009, \mn@doi [\aap] {10.1051/0004-6361/200911798}, \href
  {https://ui.adsabs.harvard.edu/abs/2009A&A...504..359J} {504, 359}

\bibitem[\protect\citeauthoryear{{Khandai}, {Di Matteo}, {Croft}, {Wilkins},
  {Feng}, {Tucker}, {DeGraf}  \& {Liu}}{{Khandai} et~al.}{2015}]{kdc15}
{Khandai} N.,  {Di Matteo} T.,  {Croft} R.,  {Wilkins} S.,  {Feng} Y.,
  {Tucker} E.,  {DeGraf} C.,   {Liu} M.-S.,  2015, \mn@doi [\mnras]
  {10.1093/mnras/stv627}, \href
  {https://ui.adsabs.harvard.edu/abs/2015MNRAS.450.1349K} {450, 1349}

\bibitem[\protect\citeauthoryear{{Khostovan}, {Sobral}, {Mobasher}, {Best},
  {Smail}, {Stott}, {Hemmati}  \& {Nayyeri}}{{Khostovan} et~al.}{2015}]{ksm15}
{Khostovan} A.~A.,  {Sobral} D.,  {Mobasher} B.,  {Best} P.~N.,  {Smail} I.,
  {Stott} J.~P.,  {Hemmati} S.,   {Nayyeri} H.,  2015, \mn@doi [\mnras]
  {10.1093/mnras/stv1474}, \href
  {https://ui.adsabs.harvard.edu/abs/2015MNRAS.452.3948K} {452, 3948}

\bibitem[\protect\citeauthoryear{{Kilbinger}}{{Kilbinger}}{2015}]{Kilbinger15}
{Kilbinger} M.,  2015, \mn@doi [Reports on Progress in Physics]
  {10.1088/0034-4885/78/8/086901}, \href
  {https://ui.adsabs.harvard.edu/abs/2015RPPh...78h6901K} {78, 086901}

\bibitem[\protect\citeauthoryear{{Krause} \& {Eifler}}{{Krause} \&
  {Eifler}}{2017}]{Krause_Tim17}
{Krause} E.,  {Eifler} T.,  2017, \mn@doi [\mnras] {10.1093/mnras/stx1261},
  \href {https://ui.adsabs.harvard.edu/abs/2017MNRAS.470.2100K} {470, 2100}

\bibitem[\protect\citeauthoryear{{Krause}, {Eifler}  \& {Blazek}}{{Krause}
  et~al.}{2016}]{Krause16_IA}
{Krause} E.,  {Eifler} T.,   {Blazek} J.,  2016, \mn@doi [\mnras]
  {10.1093/mnras/stv2615}, \href
  {https://ui.adsabs.harvard.edu/abs/2016MNRAS.456..207K} {456, 207}

\bibitem[\protect\citeauthoryear{{Laigle} et~al.,}{{Laigle}
  et~al.}{2016}]{COSMOS2015}
{Laigle} C.,  et~al., 2016, \mn@doi [\apjs] {10.3847/0067-0049/224/2/24}, \href
  {https://ui.adsabs.harvard.edu/abs/2016ApJS..224...24L} {224, 24}

\bibitem[\protect\citeauthoryear{{Laureijs} et~al.,}{{Laureijs}
  et~al.}{2011}]{Euclid_DSR}
{Laureijs} R.,  et~al., 2011, arXiv e-prints, \href
  {https://ui.adsabs.harvard.edu/abs/2011arXiv1110.3193L} {p. arXiv:1110.3193}

\bibitem[\protect\citeauthoryear{{Lin}, {Harnois-D{\'e}raps}, {Eifler},
  {Pospisil}, {Mandelbaum}, {Lee}, {Singh}  \& {LSST Dark Energy Science
  Collaboration}}{{Lin} et~al.}{2020}]{likelihood}
{Lin} C.-H.,  {Harnois-D{\'e}raps} J.,  {Eifler} T.,  {Pospisil} T.,
  {Mandelbaum} R.,  {Lee} A.~B.,  {Singh} S.,   {LSST Dark Energy Science
  Collaboration} 2020, \mn@doi [\mnras] {10.1093/mnras/staa2948}, \href
  {https://ui.adsabs.harvard.edu/abs/2020MNRAS.499.2977L} {499, 2977}

\bibitem[\protect\citeauthoryear{{Mandelbaum}}{{Mandelbaum}}{2018}]{Mandelbaum18}
{Mandelbaum} R.,  2018, \mn@doi [\araa] {10.1146/annurev-astro-081817-051928},
  \href {https://ui.adsabs.harvard.edu/abs/2018ARA&A..56..393M} {56, 393}

\bibitem[\protect\citeauthoryear{{Mandelbaum} et~al.,}{{Mandelbaum}
  et~al.}{2018}]{MLL+18}
{Mandelbaum} R.,  et~al., 2018, \mn@doi [\mnras] {10.1093/mnras/sty2420}, \href
  {https://ui.adsabs.harvard.edu/abs/2018MNRAS.481.3170M} {481, 3170}

\bibitem[\protect\citeauthoryear{{Miller} et~al.,}{{Miller}
  et~al.}{2013}]{Miller13}
{Miller} L.,  et~al., 2013, \mn@doi [\mnras] {10.1093/mnras/sts454}, \href
  {https://ui.adsabs.harvard.edu/abs/2013MNRAS.429.2858M} {429, 2858}

\bibitem[\protect\citeauthoryear{{Morales}}{{Morales}}{2006}]{Morales06}
{Morales} M.~F.,  2006, \mn@doi [\apjl] {10.1086/508614}, \href
  {https://ui.adsabs.harvard.edu/abs/2006ApJ...650L..21M} {650, L21}

\bibitem[\protect\citeauthoryear{{Outini} \& {Copin}}{{Outini} \&
  {Copin}}{2020}]{OC20}
{Outini} M.,  {Copin} Y.,  2020, \mn@doi [\aap] {10.1051/0004-6361/201936318},
  \href {https://ui.adsabs.harvard.edu/abs/2020A&A...633A..43O} {633, A43}

\bibitem[\protect\citeauthoryear{{Pillepich} et~al.,}{{Pillepich}
  et~al.}{2018}]{psn18}
{Pillepich} A.,  et~al., 2018, \mn@doi [\mnras] {10.1093/mnras/stx2656}, \href
  {https://ui.adsabs.harvard.edu/abs/2018MNRAS.473.4077P} {473, 4077}

\bibitem[\protect\citeauthoryear{{Planck Collaboration} et~al.,}{{Planck
  Collaboration} et~al.}{2020}]{Planck2018}
{Planck Collaboration} et~al., 2020, \mn@doi [\aap]
  {10.1051/0004-6361/201833910}, \href
  {https://ui.adsabs.harvard.edu/abs/2020A&A...641A...6P} {641, A6}

\bibitem[\protect\citeauthoryear{{R. S.}, {Krause}, {Huang}, {Huff}, {Xu},
  {Eifler}  \& {Everett}}{{R. S.} et~al.}{2022}]{SKH+22}
{R. S.} P.,  {Krause} E.,  {Huang} H.-J.,  {Huff} E.,  {Xu} J.,  {Eifler} T.,
  {Everett} S.,  2022, arXiv e-prints, \href
  {https://ui.adsabs.harvard.edu/abs/2022arXiv220911811S} {p. arXiv:2209.11811}

\bibitem[\protect\citeauthoryear{{Reyes}, {Mandelbaum}, {Gunn}, {Pizagno}  \&
  {Lackner}}{{Reyes} et~al.}{2011}]{R11}
{Reyes} R.,  {Mandelbaum} R.,  {Gunn} J.~E.,  {Pizagno} J.,   {Lackner} C.~N.,
  2011, \mn@doi [\mnras] {10.1111/j.1365-2966.2011.19415.x}, \href
  {https://ui.adsabs.harvard.edu/abs/2011MNRAS.417.2347R} {417, 2347}

\bibitem[\protect\citeauthoryear{{Saito}, {de la Torre}, {Ilbert}, {Dubois},
  {Yabe}  \& {Coupon}}{{Saito} et~al.}{2020}]{sti20}
{Saito} S.,  {de la Torre} S.,  {Ilbert} O.,  {Dubois} C.,  {Yabe} K.,
  {Coupon} J.,  2020, \mn@doi [\mnras] {10.1093/mnras/staa727}, \href
  {https://ui.adsabs.harvard.edu/abs/2020MNRAS.494..199S} {494, 199}

\bibitem[\protect\citeauthoryear{{Schaye} et~al.,}{{Schaye}
  et~al.}{2010}]{sdb10}
{Schaye} J.,  et~al., 2010, \mn@doi [\mnras]
  {10.1111/j.1365-2966.2009.16029.x}, \href
  {https://ui.adsabs.harvard.edu/abs/2010MNRAS.402.1536S} {402, 1536}

\bibitem[\protect\citeauthoryear{{Schaye} et~al.,}{{Schaye}
  et~al.}{2015}]{scb15}
{Schaye} J.,  et~al., 2015, \mn@doi [\mnras] {10.1093/mnras/stu2058}, \href
  {https://ui.adsabs.harvard.edu/abs/2015MNRAS.446..521S} {446, 521}

\bibitem[\protect\citeauthoryear{{Secco} et~al.,}{{Secco}
  et~al.}{2021}]{DES_Y3_WL2}
{Secco} L.~F.,  et~al., 2021, arXiv e-prints, \href
  {https://ui.adsabs.harvard.edu/abs/2021arXiv210513544S} {p. arXiv:2105.13544}

\bibitem[\protect\citeauthoryear{{Seitz} \& {Schneider}}{{Seitz} \&
  {Schneider}}{1997}]{SS97}
{Seitz} C.,  {Schneider} P.,  1997, \aap, \href
  {https://ui.adsabs.harvard.edu/abs/1997A&A...318..687S} {318, 687}

\bibitem[\protect\citeauthoryear{{Sobral}, {Smail}, {Best}, {Geach}, {Matsuda},
  {Stott}, {Cirasuolo}  \& {Kurk}}{{Sobral} et~al.}{2013}]{Sobral13}
{Sobral} D.,  {Smail} I.,  {Best} P.~N.,  {Geach} J.~E.,  {Matsuda} Y.,
  {Stott} J.~P.,  {Cirasuolo} M.,   {Kurk} J.,  2013, \mn@doi [\mnras]
  {10.1093/mnras/sts096}, \href
  {https://ui.adsabs.harvard.edu/abs/2013MNRAS.428.1128S} {428, 1128}

\bibitem[\protect\citeauthoryear{{Spergel} et~al.,}{{Spergel}
  et~al.}{2015}]{WFIRST_AFTA}
{Spergel} D.,  et~al., 2015, arXiv e-prints, \href
  {https://ui.adsabs.harvard.edu/abs/2015arXiv150303757S} {p. arXiv:1503.03757}

\bibitem[\protect\citeauthoryear{{Square Kilometre Array Cosmology Science
  Working Group} et~al.,}{{Square Kilometre Array Cosmology Science Working
  Group} et~al.}{2020}]{SKA_RedPaper}
{Square Kilometre Array Cosmology Science Working Group} et~al., 2020, \mn@doi
  [\pasa] {10.1017/pasa.2019.51}, \href
  {https://ui.adsabs.harvard.edu/abs/2020PASA...37....7S} {37, e007}

\bibitem[\protect\citeauthoryear{{Stil}, {Krause}, {Beck}  \& {Taylor}}{{Stil}
  et~al.}{2009}]{SKBT_09}
{Stil} J.~M.,  {Krause} M.,  {Beck} R.,   {Taylor} A.~R.,  2009, \mn@doi [\apj]
  {10.1088/0004-637X/693/2/1392}, \href
  {https://ui.adsabs.harvard.edu/abs/2009ApJ...693.1392S} {693, 1392}

\bibitem[\protect\citeauthoryear{{Sun} \& {Reich}}{{Sun} \&
  {Reich}}{2012}]{SR12}
{Sun} X.~H.,  {Reich} W.,  2012, \mn@doi [\aap] {10.1051/0004-6361/201218802},
  \href {https://ui.adsabs.harvard.edu/abs/2012A&A...543A.127S} {543, A127}

\bibitem[\protect\citeauthoryear{{Takada} \& {Hu}}{{Takada} \&
  {Hu}}{2013}]{TH13}
{Takada} M.,  {Hu} W.,  2013, \mn@doi [\prd] {10.1103/PhysRevD.87.123504},
  \href {https://ui.adsabs.harvard.edu/abs/2013PhRvD..87l3504T} {87, 123504}

\bibitem[\protect\citeauthoryear{{Takada} \& {Jain}}{{Takada} \&
  {Jain}}{2009}]{tj09}
{Takada} M.,  {Jain} B.,  2009, \mn@doi [\mnras]
  {10.1111/j.1365-2966.2009.14504.x}, \href
  {https://ui.adsabs.harvard.edu/abs/2009MNRAS.395.2065T} {395, 2065}

\bibitem[\protect\citeauthoryear{{The LSST Dark Energy Science Collaboration}
  et~al.,}{{The LSST Dark Energy Science Collaboration}
  et~al.}{2018}]{LSST_SRD18}
{The LSST Dark Energy Science Collaboration} et~al., 2018, arXiv e-prints,
  \href {https://ui.adsabs.harvard.edu/abs/2018arXiv180901669T} {p.
  arXiv:1809.01669}

\bibitem[\protect\citeauthoryear{{Tully} \& {Fisher}}{{Tully} \&
  {Fisher}}{1977}]{TF1997}
{Tully} R.~B.,  {Fisher} J.~R.,  1977, \aap, \href
  {https://ui.adsabs.harvard.edu/abs/1977A&A....54..661T} {500, 105}

\bibitem[\protect\citeauthoryear{{Weinberg}, {Mortonson}, {Eisenstein},
  {Hirata}, {Riess}  \& {Rozo}}{{Weinberg} et~al.}{2013}]{WMM13}
{Weinberg} D.~H.,  {Mortonson} M.~J.,  {Eisenstein} D.~J.,  {Hirata} C.,
  {Riess} A.~G.,   {Rozo} E.,  2013, \mn@doi [\physrep]
  {10.1016/j.physrep.2013.05.001}, \href
  {https://ui.adsabs.harvard.edu/abs/2013PhR...530...87W} {530, 87}

\bibitem[\protect\citeauthoryear{{Weinberger} et~al.,}{{Weinberger}
  et~al.}{2018}]{wsp18}
{Weinberger} R.,  et~al., 2018, \mn@doi [\mnras] {10.1093/mnras/sty1733}, \href
  {https://ui.adsabs.harvard.edu/abs/2018MNRAS.479.4056W} {479, 4056}

\bibitem[\protect\citeauthoryear{{Whittaker}, {Brown}  \& {Battye}}{{Whittaker}
  et~al.}{2015}]{WBB_15}
{Whittaker} L.,  {Brown} M.~L.,   {Battye} R.~A.,  2015, \mn@doi [\mnras]
  {10.1093/mnras/stv993}, \href
  {https://ui.adsabs.harvard.edu/abs/2015MNRAS.451..383W} {451, 383}

\bibitem[\protect\citeauthoryear{Wittman \& Self}{Wittman \& Self}{2021}]{WS19}
Wittman D.,  Self M.,  2021, \mn@doi [\apj] {10.3847/1538-4357/abd548}, \href
  {https://ui.adsabs.harvard.edu/abs/2019arXiv190504404W} {908, 34}

\bibitem[\protect\citeauthoryear{{Zhai}, {Benson}, {Wang}, {Yepes}  \&
  {Chuang}}{{Zhai} et~al.}{2019}]{zbw19}
{Zhai} Z.,  {Benson} A.,  {Wang} Y.,  {Yepes} G.,   {Chuang} C.-H.,  2019,
  \mn@doi [\mnras] {10.1093/mnras/stz2844}, \href
  {https://ui.adsabs.harvard.edu/abs/2019MNRAS.490.3667Z} {490, 3667}

\bibitem[\protect\citeauthoryear{{Zuntz} et~al.,}{{Zuntz}
  et~al.}{2018}]{ZSS+18}
{Zuntz} J.,  et~al., 2018, \mn@doi [\mnras] {10.1093/mnras/sty2219}, \href
  {https://ui.adsabs.harvard.edu/abs/2018MNRAS.481.1149Z} {481, 1149}

\bibitem[\protect\citeauthoryear{{de Burgh-Day}, {Taylor}, {Webster}  \&
  {Hopkins}}{{de Burgh-Day} et~al.}{2015a}]{DSM_15b}
{de Burgh-Day} C.~O.,  {Taylor} E.~N.,  {Webster} R.~L.,   {Hopkins} A.~M.,
  2015a, \mn@doi [\pasa] {10.1017/pasa.2015.39}, \href
  {https://ui.adsabs.harvard.edu/abs/2015PASA...32...40D} {32, e040}

\bibitem[\protect\citeauthoryear{{de Burgh-Day}, {Taylor}, {Webster}  \&
  {Hopkins}}{{de Burgh-Day} et~al.}{2015b}]{DSM_15a}
{de Burgh-Day} C.~O.,  {Taylor} E.~N.,  {Webster} R.~L.,   {Hopkins} A.~M.,
  2015b, \mn@doi [\mnras] {10.1093/mnras/stv1083}, \href
  {https://ui.adsabs.harvard.edu/abs/2015MNRAS.451.2161D} {451, 2161}

\bibitem[\protect\citeauthoryear{{van Daalen}, {Schaye}, {Booth}  \& {Dalla
  Vecchia}}{{van Daalen} et~al.}{2011}]{dsb11}
{van Daalen} M.~P.,  {Schaye} J.,  {Booth} C.~M.,   {Dalla Vecchia} C.,  2011,
  \mn@doi [\mnras] {10.1111/j.1365-2966.2011.18981.x}, \href
  {http://adsabs.harvard.edu/abs/2011MNRAS.415.3649V} {415, 3649}

\bibitem[\protect\citeauthoryear{Übler et~al.}{Übler
  et~al.}{2017}]{Ubler:2017pik}
Übler H.,  et~al., 2017, \mn@doi [Astrophys. J.] {10.3847/1538-4357/aa7558},
  842, 121

\makeatother
\end{thebibliography}

%%%%%%%%%%%%%%%%%%%%%%%%%%%%%%%%%%%%%%%%%%%%%%%%%%

%%%%%%%%%%%%%%%%% APPENDICES %%%%%%%%%%%%%%%%%%%%%

\appendix

\section{Derivation of Kinematic Lensing}
\label{sec:appd_a}

In this section, we detail the derivation of equations~(\ref{eqn:shape_result})\&(\ref{eqn:vminor}) under more general conditions. We assume a general coordinate system, which is shown in Fig.~\ref{fig:shear_illus_1}:

\begin{figure*}
	\includegraphics[scale=0.5]{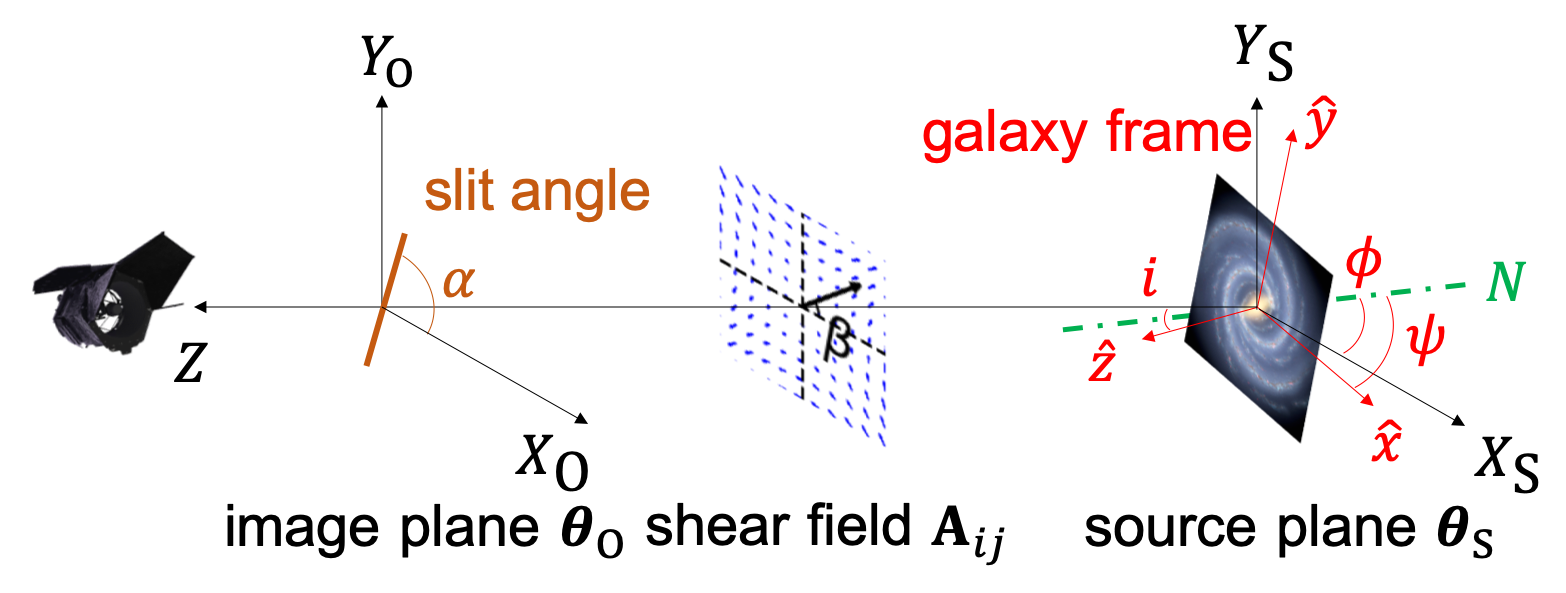}
    \caption{Definition of coordinate systems. The galaxy intrinsic frame is denoted as galaxy frame in red. The $\hat{z}$-axis is defined by the direction of the angular momentum of the disc. The galaxy frame and the comoving frame at the source galaxy, the source plane ($X_{\mathrm{S}}$-$Y_{\mathrm{S}}$-$Z$), are connected by the Euler angles $(\phi, i, \psi)$ defined in the plot. Here $\phi$ and $i$ work as the position and inclination angles. The line of nodes is shown in green. The image plane ($X_{\mathrm{O}}$-$Y_{\mathrm{O}}$-$Z$) is the observed frame after being lensed. The position angle $\beta$ of the shear field is defined as the direction along which the shear matrix is diagonal. In the presence of dispersing elements, the cross-dispersion direction, or the slit angle, is denoted as $\alpha$.}
    \label{fig:shear_illus_1}
\end{figure*}

\begin{itemize}
    \item Galaxy frame ($\hat{x}$-$\hat{y}$-$\hat{z}$): the frame aligned with the source galaxy, where the $\hat{z}$-axis is aligned with the angular momentum of the disc.
    
    \item Source plane ($X_\mrm{S}$-$Y_\mrm{S}$): the observer frame at the redshift of source galaxy. Generally $X_\mrm{S}$-$Y_\mrm{S}$ can be arbitrary but a nature choice of $X_{\mrm{S}}$-$Y_{\mrm{S}}$ (and also $X_\mrm{O}$-$Y_\mrm{O}$) are RA and DEC accordingly. For convenience, we define $Z$-axis as the inverse of LoS direction. The galaxy frame and source plane are connected by the Euler angles $(\phi, i, \psi)$. We will also refer to $i$ as the inclination angle and $\phi$ the position angle. Vectors in the source plane are denoted as $\bm{\theta}_{\mrm{S}}$.
    
    \item Image plane ($X_\mrm{O}$-$Y_\mrm{O}$): the observer frame at $z=0$. Vectors in the image plane are denoted as $\bm{\theta}_{\mrm{O}}$.
\end{itemize}
Accordingly, we define the distortion matrix as 
\begin{equation}
\label{eqn:shear_mat_def}
    \mat{A}
    \equiv \frac{\partial\bm{\theta}_{\mrm{S}}}{\partial\bm{\theta}_{\mrm{O}}}=
    \begin{pmatrix}
    1-\kappa-\gamma_1 & -\gamma_2 \\
    -\gamma_2 & 1-\kappa+\gamma_1
    \end{pmatrix},
\end{equation}
such that $\bm{\theta}_{\mrm{S}}=\mat{A}\bm{\cdot}\bm{\theta}_\mrm{O}$.

%%% angles & axes
We also define the following angles/axes:
\begin{itemize}
    \item Position angle (PA) of photometric major/minor axis: we refer to the major/minor axis of the photometric image of the source galaxy as photometric major/minor axis. The PA of the major axis is denoted as $\phi$ in the source plane and $\phi^\prime$ in the image plane. The PAs of minor axes are $\phi+\pi/2$ and $\phi^\prime+\pi/2$ correspondingly. Note that under such convention, the $g_{+/\times}$ defined in Sect.~\ref{sec:KL} are $g_+=g_1\cos{2\phi^\prime}+g_2\sin{2\phi^\prime}$ and $g_\times=g_2\cos{2\phi^\prime}-g_1\sin{2\phi^\prime}$.
    
    \item PA of kinematic major/minor axis: we refer to the kinematic major/minor axis as the axis along which the LoS velocity has the maximum/minimum gradient. We also call the kinematic minor axis as the zero-LoS-velocity direction. In the source plane, the kinematic major/minor axis is aligned with the photometric major/minor axis.
    
    \item PA of shear field: the PA of shear field, $\beta$, is defined such that, when rotating the $X_\mrm{O}$-$Y_\mrm{O}$ frame by $\beta$, $\gamma_2=0$ in the new frame. Thus $(\gamma_1,\gamma_2)$ could also be written as $(\gamma\cos{2\beta},\gamma\sin{2\beta})$, where $\gamma=\sqrt{\gamma_1^2+\gamma_2^2}$. Or expressed by complex number, $\bm{\gamma}=\gamma_1+i\gamma_2=\gamma e^{2i\beta}$
    
    \item Slit angle/dispersion direction: In slit spectroscopy, we refer to the PA of the slit as $\alpha$ and to the PA of the dispersion direction as $\alpha-\pi/2$. In slitless spectroscopy, we keep the term dispersion direction, while using the slit angle and the PA of cross-dispersion direction interchangeably. 
\end{itemize}

The 2nd moment defined in equation~(\ref{eqn:Iij_def}) plays an essential part in connecting the shape measurements in the image plane $I_{ij}$ and the source plane $I_{ij}^\mrm{int}$: 
\begin{equation}
\label{eqn:Iij_image_source}
\begin{split}
I_{ij}&
=\int\mrm{d}\bm{\theta}_\mrm{O}I(\bm{\theta}_\mrm{O})\theta_{i,\mrm{O}}\theta_{j,\mrm{O}}
=\int\mrm{d}\bm{\theta}_\mrm{O} I(\bm{\theta}_\mrm{O})[\bm{\theta}_\mrm{O}\bm{\otimes}\bm{\theta}_\mrm{O}^\mrm{T}]_{ij}\\
&=\int\mrm{d}\bm{\theta}_\mrm{S} \begin{vmatrix}\frac{\partial \theta_\mrm{O}}{\partial \theta_\mrm{S}}\end{vmatrix} I(\bm{\theta}_\mrm{S}) [\mat{A}^{-1} \bm{\cdot} (\bm{\theta}_\mrm{S} \bm{\otimes} \bm{\theta}_\mrm{S}^\mrm{T}) \bm{\cdot} (\mat{A}^{-1})^\mrm{T}]_{ij} \\
&= [\mat{A}^{-1} \bm{\cdot}(\int\mrm{d}\bm{\theta}_\mrm{S} I(\bm{\theta}_\mrm{S})\,\bm{\theta}_\mrm{S} \bm{\otimes} \bm{\theta}_\mrm{S}^\mrm{T}) \bm{\cdot} (\mat{A}^{-1})^\mrm{T}]_{ij}/\begin{vmatrix}\frac{\partial \theta_\mrm{S}}{\partial \theta_\mrm{O}}\end{vmatrix} \\
    &= [\mat{A}^{-1}\bm{\cdot} I^\mrm{int}\bm{\cdot}(\mat{A}^{-1})^\mrm{T}]_{ij}/\mrm{det}(\mat{A}),
\end{split}
\end{equation}
where we utilize the property that weak lensing conserves the specific intensity $I(\bm{\theta}_\mrm{O})=I(\bm{\theta}_\mrm{S})$, and assume the shear matrix $\mat{A}$ is constant within the galaxy image. 
Switching $I_{ij}$ and $I_{ij}^\mrm{int}$ and replace $\mat{A}$ by $\mat{A}^{-1}$ gives the inverse relation
\begin{equation}
    \label{eqn:Iij_source_image}
    I_{ij}^\mrm{int} = [\mat{A}\bm{\cdot} I\bm{\cdot}\mat{A}^\mrm{T}]_{ij}\,\mrm{det}(\mat{A}).
\end{equation}
Substituting equation~(\ref{eqn:Iij_image_source}) into equation~(\ref{eqn:quadratic_estimators_def}) gives~\citep{BS01_review}
\begin{equation}
\label{eqn:estimators_image_to_source}
    \hat{\bm{\epsilon}} = \frac{\hat{\bm{\epsilon}}^\mrm{int}+\bm{g}}{1+\bm{g}^*\hat{\bm{\epsilon}}^\mathrm{int}},
\end{equation}
where $\bm{g}$ is the complex reduced shear field and $\bm{g}^*$ is its complex conjugate. 
Note that we only take the $|\bm{g}|<1$ part of the expression in \cite{BS01_review}. 

To get the relation between $e_\mrm{int}$ and $e_\mrm{obs}$, consider the inverse relation of equation~(\ref{eqn:estimators_image_to_source}):
\begin{equation}
\label{eqn:estimators_source_to_image}
    \hat{\bm{\epsilon}}^\mrm{int} = \frac{\hat{\bm{\epsilon}}-\bm{g}}{1-\bm{g}^*\hat{\bm{\epsilon}}},
\end{equation}
the expanded version is
\begin{equation}
\label{eqn:estimators_source_to_image_long}
    \begin{aligned}
        \hat{\epsilon}_1^\mrm{int} &= \frac{\hat{\epsilon}_1-(1+e_\mrm{obs}^2)g_1+(g_1^2-g_2^2)\hat{\epsilon}_1+2g_1g_2\hat{\epsilon}_2}{1-2(g_1\hat{\epsilon}_1+g_2\hat{\epsilon}_2)+e_\mrm{obs}^2g^2}\\
        \hat{\epsilon}_2^\mrm{int} &= \frac{\hat{\epsilon}_2-(1+e_\mrm{obs}^2)g_2+2g_1g_2\hat{\epsilon}_1-(g_1^2-g_2^2)\hat{\epsilon}_2}{1-2(g_1\hat{\epsilon}_1+g_2\hat{\epsilon}_2)+e_\mrm{obs}^2g^2}
    \end{aligned}
\end{equation}
multiplying equation~(\ref{eqn:estimators_source_to_image}) by $(\hat{\bm{\epsilon}}^\mrm{int})^*$ gives
\begin{equation}
    |(\hat{\bm{\epsilon}}^\mrm{int})^2||1-\bm{g}^*\hat{\bm{\epsilon}}|^2=|\hat{\bm{\epsilon}}-\bm{g}|^2
\end{equation}
Simply substituting $\bm{g}$ and $\hat{\bm{\epsilon}}$ would give equation~(\ref{eqn:shape_result}).

Next, let us consider the distortion on LoS velocity in the image plane $v_\mrm{LoS}^\prime(\bm{\theta}_\mrm{O})$.
As we discussed in equation~(\ref{eqn:formal_velocity}), the observed LoS velocity is connected with the one in the source plane via the distortion matrix mapping. To model $v_\mrm{LoS}(\bm{\theta}_\mrm{S})$, we first consider the velocity in the galaxy frame. Note that the galaxy disc is invariant under rotation around the $\hat{z}$-axis, thus we can tune $\psi$ such that, in the galaxy frame, its position is $\bm{r}=(r,\,0,\,0)^\mrm{T}$. The velocity vector is $\bm{v}=(0,\,V(r),\,0)^\mrm{T}$, where $V(r)$ is the rotation curve model. A widely used one is
\begin{equation}
\label{eqn:rotation_curve_model}
    V(r) = V_0+\frac{V_\mrm{circ}}{\pi/2}\mrm{tan}^{-1}(\frac{r-r_0}{r_{\mrm{v}\textit{-}\mrm{scale}}}),
\end{equation}
where $V_0$ is the systemic velocity including both the Hubble flow and the peculiar velocity of the galaxy, $r_0$ and $r_{\mrm{v}\textit{-}\mrm{scale}}$ are parameters controlling the turnover radius and the slope of the rotation curve, $V_\mrm{circ}$ is the asymptotic velocity when $r$ is much larger than the scale radius, or the plateau velocity. In this work, $V_\mrm{circ}$ is equivalent to $v_\mrm{TF}$. 

To project $\bm{r}$ and $\bm{v}$ into the source plane, consider the rotation matrix $\mat{R}(\phi,\,i,\,\psi)$,
\newpage
\begin{widetext}
    \begin{equation}
    \label{eqn:Euler_angle_matrix}
    \mat{R}(\phi,\,i,\,\psi)\equiv
    \begin{pmatrix}
        \cos{\phi}\cos{\psi} - \cos{i}\sin{\phi}\sin{\psi} & -\cos{\phi}\sin{\psi}-\cos{i}\sin{\phi}\cos{\psi} & \sin{i}\sin{\phi} \\
        \sin{\phi}\cos{\psi} + \cos{i}\cos{\phi}\sin{\psi} & -\sin{\phi}\sin{\psi}+\cos{i}\cos{\phi}\cos{\psi} & -\sin{i}\cos{\phi} \\
        \sin{i}\sin{\psi} & \sin{i}\cos{\psi} & \cos{i}\\
    \end{pmatrix},
\end{equation}
\end{widetext}
as a function of the Euler angles shown in Fig.~\ref{fig:shear_illus_1}. In the source plane, the 3D position vector $\bm{r}_\mrm{S}$ and velocity vector $\bm{v}_\mrm{S}$ are given by

\begin{equation}
    \label{eqn:3D_rs}
    \begin{aligned}
    \bm{r}_\mrm{S}&=\mat{R}(\phi,\,i,\,\psi)\bm{\cdot}\bm{r}\\
    &=\begin{pmatrix}
        r\cos{\phi}\cos{\psi}-r\cos{i}\sin{\phi}\sin{\psi}\\
        r\sin{\phi}\cos{\psi}+r\cos{i}\cos{\phi}\sin{\psi}\\
        r\sin{i}\sin{\psi}\\
    \end{pmatrix},
    \end{aligned}
\end{equation}

\begin{equation}
    \label{eqn:3D_vs}
    \begin{aligned}
    \bm{v}_\mrm{S}&=\mat{R}(\phi,\,i,\,\psi)\bm{\cdot}\bm{v}\\
    &=\begin{pmatrix}
        -V(r)\cos{\phi}\sin{\psi}-V(r)\cos{i}\sin{\phi}\cos{\psi}\\
        -V(r)\sin{\phi}\sin{\psi}+V(r)\cos{i}\cos{\phi}\cos{\psi}\\
        V(r)\sin{i}\cos{\psi}\\
    \end{pmatrix}.
    \end{aligned}
\end{equation}
%\end{widetext}

The projected position $\bm{\theta}_\mrm{S}$ and LoS velocity $v_\mrm{LoS}$ are hence
\begin{equation}
    \label{eqn:galaxy_frame_source_plane_position}
    \bm{\theta}_\mrm{S}=\begin{pmatrix}
    r\cos{\phi}\cos{\psi}-r\cos{i}\sin{\phi}\sin{\psi}\\
    r\sin{\phi}\cos{\psi}+r\cos{i}\cos{\phi}\sin{\psi}\\
    \end{pmatrix}=\mat{A}\bm{\cdot}\bm{\theta}_\mrm{O},
\end{equation}
\begin{equation}
    \label{eqn:galaxy_frame_source_plane_velocity}
    v_\mrm{LoS}=-V(r)\sin{i}\cos{\psi}=v_\mrm{LoS}(\bm{\theta}_\mrm{S})=v_\mrm{LoS}^\prime(\bm{\theta}_\mrm{O}).
\end{equation}
Note that the minus sign in equation~(\ref{eqn:galaxy_frame_source_plane_velocity}) is applied to change $Z$-axis outward along LoS.

Besides $i$ which can be inferred from the TFR, we still need $\psi$ and $r$ to calculate $v_\mrm{LoS}$ from $\bm{\theta}_\mrm{S}$, given a rotation curve model. Using that $\phi=\phi^\prime-(\phi^\prime-\phi)$, both $\phi^\prime$ and $\Delta\phi\equiv\phi^\prime-\phi$ can be calculated from observations: $\phi^\prime$ is directly estimated from the image plane photometry, while $\Delta\phi$ is a little bit cumbersome. Consider 3D vector cross product $(\epsilon_1,\,\epsilon_2,\,0)^\mrm{T}\bm{\times} 
(\epsilon_1^\mrm{int},\,\epsilon_2^\mrm{int},\,0)^\mrm{T}=-e_\mrm{obs}e_\mrm{int}\sin{2\Delta\phi}\,\bm{Z}$, where $\bm{Z}$ is the unit vector along $Z$-axis, substituting $\bm{\epsilon}^\mrm{int}$ with equation~(\ref{eqn:estimators_source_to_image_long}), we have 
\begin{equation}
\begin{split}
    \sin{2\Delta\phi}=&\frac{(1+e_\mrm{obs}^2)(g_2\hat{\epsilon}_1-g_1\hat{\epsilon}_2)}{e_\mrm{int}e_\mrm{obs}}\\
    &+\frac{2e_\mrm{obs}}{e_\mrm{int}}[g_1g_2(\hat{\epsilon}_1^2-\hat{\epsilon}_2^2)-(g_1^2-g_2^2)\hat{\epsilon}_1\hat{\epsilon}_2]\\
    &+\bigo{g^3}.
\end{split}
\end{equation}
Note that the first term of $\sin{2\Delta\phi}\propto g^1e_\mrm{int}^{-1}$, thus when $e_\mrm{int}\approx g$, the change of PA is a finite quantity and arcsin function is needed to calculate $\Delta\phi$. The second term of $\sin{2\Delta\phi}\propto g^2e_\mrm{obs}^2$, thus is safe to ignore. In case of $e_\mrm{int}\gg g$, $\Delta\phi$ can be further simplified to
\begin{equation}
    \Delta\phi\approx \frac{(1+e_\mrm{obs}^2)(g_2\hat{\epsilon}_1-g_1\hat{\epsilon}_2)}{2e_\mrm{int}e_\mrm{obs}}.
\end{equation}
However, as we see in Fig.~\ref{fig:rand_sanple}, there is still non-negligible number of galaxies of which $e_\mrm{int}<0.1$ in a general galaxy sample, so this approximation should be used with caution. Generally, we can solve for $\phi$ and write down the rotation matrix
\begin{equation}
    \mat{R}^{-1}(\phi^\prime, \Delta\phi)=\begin{pmatrix}
        \cos{(\phi^\prime-\Delta\phi)} & \sin{(\phi^\prime-\Delta\phi)}\\
        -\sin{(\phi^\prime-\Delta\phi)} & \cos{(\phi^\prime-\Delta\phi)}\\
    \end{pmatrix},
\end{equation}
thus
\begin{equation}
    \mat{R}^{-1}(\phi^\prime, \Delta\phi)\bm{\cdot\theta}_\mrm{S}=\begin{pmatrix}
        r\cos{\psi}\\
        r\cos{i}\sin{\psi}
    \end{pmatrix}.
\end{equation}
As a result, $r$ and $\cos{\psi}$ can be written as
\begin{equation}
\label{eqn:r_theta_O}
    r(\bm{\theta}_\mrm{O}) = \left[\bm{\theta}(\bm{\theta}_\mrm{O})^\mrm{T}\bm{\cdot}\begin{pmatrix}
        1 & 0\\
        0 & 1/\cos{^2i}\\
    \end{pmatrix}\bm{\cdot\theta}(\bm{\theta}_\mrm{O})\right]^{1/2},
\end{equation}
\begin{equation}
\label{eqn:cospsi_theta_O}
    \cos{\psi}=\left[(1,0)\bm{\cdot\theta}(\bm{\theta}_\mrm{O})\right]/r(\bm{\theta}_\mrm{O}),
\end{equation}
where
\begin{equation}
\label{eqn:theta_gal}
    \bm{\theta}(\bm{\theta}_\mrm{O})=\mat{R}^{-1}(\phi^\prime,\Delta\phi)\bm{\cdot }\mat{A}\bm{\cdot\theta}_\mrm{O}.
\end{equation}
Substituting equation~(\ref{eqn:r_theta_O})\&(\ref{eqn:cospsi_theta_O}) into equation~(\ref{eqn:galaxy_frame_source_plane_velocity}), we get the general expression of the LoS velocity map in the image plane, which is a function of ($\bm{\theta}_\mrm{O}$, $\bm{g}$, $i$, $\phi$, $v_\mrm{TF}$).

For reference, we summarize some of the relevant relations given previously:
\begin{equation}
    \label{eqn:eqns_summary}
    \begin{aligned}
    \hat{\bm{\epsilon}}&=\frac{e_\mrm{int}e^{2i\phi} +\bm{g}}{1+\bm{g}^*e_\mrm{int}e^{2i\phi}},\\
    e_{\mrm{int}}&=\frac{1-\sqrt{1-(1-q_z^2)\,\sin{^2i}}}{1+\sqrt{1-(1-q_z^2)\,\sin{^2i}}},\\
    \mrm{log}_\mathrm{10}(v_{\mrm{TF}})&=a + b\, (M_\mrm{B}-M_\mrm{p}),\\
    v_1^\prime&=v_\mrm{LoS}^\prime(\bm{\theta}_1,\,\bm{g},\,i,\,\phi,\,v_\mrm{TF}),\\
    v_2^\prime&=v_\mrm{LoS}^\prime(\bm{\theta}_2,\,\bm{g},\,i,\,\phi,\,v_\mrm{TF}),\\
    \end{aligned}  
\end{equation}
where $v_1^\prime$ and $v_2^\prime$ are LoS velocities measured along two arbitrary positions $\bm{\theta}_1$ and $\bm{\theta}_2$, and are known quantities. TFR parameters, $a$ and $b$ are self-calibrated or calibrated with external datasets. The unknown variables in equation~(\ref{eqn:eqns_summary}) are $g_1$, $g_2$, $e_\mrm{int}$, $\phi$, $v_\mrm{TF}$ and $i$, thus equation~(\ref{eqn:eqns_summary}) forms a closed equation set with 6 parameters and 6 equations, and we can solve for $g_{1/2}$ and $\epsilon_{1,2}^\mrm{int}$ for general measurements.

Consider two special positions, i.e. the photometric major and minor axis, $\bm{\theta}_\mrm{major}$ = $r$($\cos{\phi^\prime}$, $\sin{\phi^\prime}$)$^\mrm{T}$ and $\bm{\theta}_\mrm{minor}$ = $r$($-\sin{\phi^\prime}$, $\cos{\phi^\prime}$)$^\mrm{T}$, substituting them into equation~(\ref{eqn:theta_gal}), we have 

\begin{equation}
    \label{eqn:theta_major/minor_gal}
    \begin{aligned}
        \bm{\theta}(\bm{\theta}_\mrm{major})&=r
        \begin{pmatrix}
        \cos{\Delta\phi}-\Re(\bm{g}e^{-i(2\phi^\prime-\Delta\phi)})\\
        \sin{\Delta\phi}-\Im(\bm{g}e^{-i(2\phi^\prime-\Delta\phi)})\\
        \end{pmatrix},\\
        \bm{\theta}(\bm{\theta}_\mrm{minor})&=r
        \begin{pmatrix}
        -\sin{\Delta\phi}-\Im(\bm{g}e^{-i(2\phi^\prime-\Delta\phi)})\\
        \cos{\Delta\phi}+\Re(\bm{g}e^{-i(2\phi^\prime-\Delta\phi)})\\
        \end{pmatrix},\\
    \end{aligned}
\end{equation}
where $\Re(z)$ and $\Im(z)$ are real and imaginary parts of $z$.
For the limiting case considered in Sect.~\ref{sec:KL_kine}, i.e.
$\Delta\phi\ll 1$, equation~(\ref{eqn:theta_major/minor_gal}) reduce to
\begin{equation}
\begin{split}
    \bm{\theta}(\bm{\theta}_\mrm{major})&=\frac{r}{e_\mrm{obs}}\begin{pmatrix}
        e_\mrm{obs}-(g_1\hat{\epsilon}_1+g_2\hat{\epsilon}_2)\\
        (\frac{1+e_\mrm{obs}^2}{2e_\mrm{int}}-1)(g_2\hat{\epsilon}_1-g_1\hat{\epsilon}_2)\\
    \end{pmatrix}+\bigo{g^2}\\
    \bm{\theta}(\bm{\theta}_\mrm{minor})&=\frac{r}{e_\mrm{obs}}\begin{pmatrix}
    (\frac{1+e_\mrm{obs}^2}{2e_\mrm{int}}+1)(g_1\hat{\epsilon}_2-g_2\hat{\epsilon}_1)\\
        e_\mrm{obs}+(g_1\hat{\epsilon}_1+g_2\hat{\epsilon}_2)\\
    \end{pmatrix}+\bigo{g^2}\\
\end{split}
\end{equation}
The resulting $\cos{\psi}$ are
\begin{equation}
\label{eqn:cospsi_special_case}
\begin{split}
    \cos{\psi_\mrm{major}}&=1+\bigo{g^2}\\
    \cos{\psi_\mrm{minor}}&=-\cos{i}\frac{g_2\hat{\epsilon}_1-g_1\hat{\epsilon}_2}{e_\mrm{obs}}\left(\frac{1+e_\mrm{obs}^2}{2e_\mrm{int}}+1\right)+\bigo{g^2}\\
\end{split}
\end{equation}
Substitute equation~(\ref{eqn:cospsi_special_case}) into equation~(\ref{eqn:galaxy_frame_source_plane_velocity}) and assume a constant rotation velocity $v_\mrm{TF}$, we get equation~(\ref{eqn:vminor}).

\section{Shape Noise Estimation}
\label{sec:shape_noise_estimation_details}

\begin{figure}
    \centering
    \includegraphics[width=\linewidth]{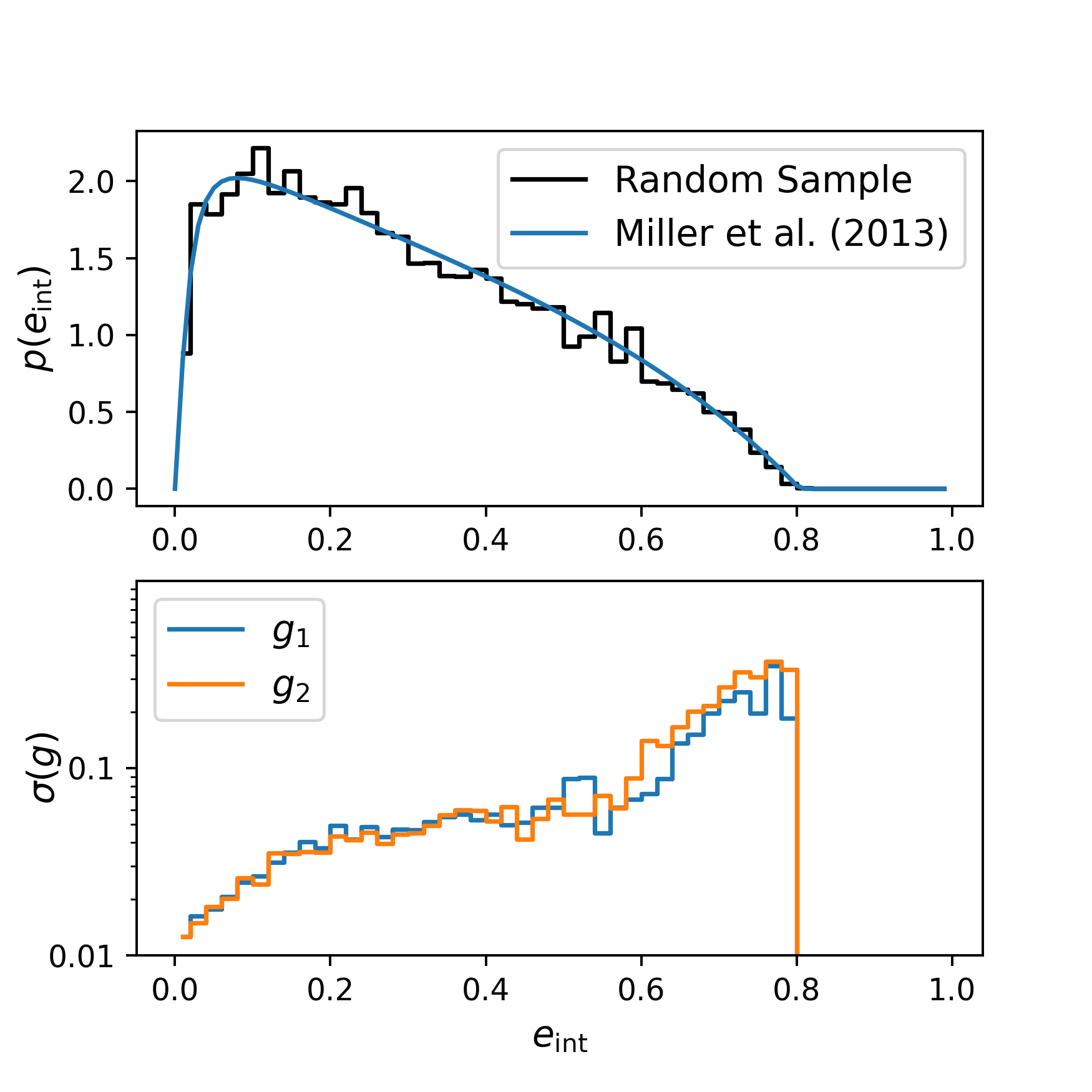}
    \caption{\textbf{Upper panel}: We show the distribution on $e_\mrm{int}$ of our random sample by the black solid histogram. As a comparison, we also show the analytic fitting function~\citep{Miller13} as the blue solid line. \textbf{Lower Panel}: Here we show the distribution of shape noise (blue for the $g_1$ noise and orange for the $g_2$ noise) binned by $e_\mrm{int}$. Here we assume velocity measurement error $\sigma_v=15$ km$\,$s$^{-1}$.}
    \label{fig:rand_sanple}
\end{figure}

In this appendix, we elaborate on our toy model shape noise calculation. As briefly discussed in Sect.~\ref{sec:shape_noise}, the calculation is divided into following steps:

(1) We generate a random ensemble of model parameters $\bm{X}_\mrm{model}$=($e_\mrm{int}$, $\phi$, $\bm{g}$, $v_\mrm{TF}$), with $e_\mrm{int}$ following the distribution in \cite{Miller13}. (see %the upper panel in
Fig.~\ref{fig:rand_sanple})

\begin{equation}
\label{eqn:eint_prior}
    p(e_\mrm{int})=\frac{Ae(1-\mrm{exp}[(e-e_\mrm{max})/a])}{(1+e)\sqrt{e^2+e_0^2}}\,,
\end{equation}
and we take the fiducial parameters the same as their work ($e_\mrm{max}$, $a$, $e_0$)=(0.804, 0.2539, 0.0256), and the amplitude $A=2.432$. $\phi$ follows a uniform distribution among [$-\pi/2$, $\pi/2$), and $v_\mrm{TF}$ follows $\mathcal{N}$(200, 20) km$\,$s$^{-1}$. $\bm{g}$ is fixed to the fiducial value, in our case is $0.05\times\mrm{exp}(i\pi/4)$. We fix $q_z=0.1$ such that the maximum intrinsic ellipticity is smaller than $e_\mrm{max}$. 
We note that the distribution from equation~\ref{eqn:eint_prior} is biased towards low inclination angles or face-on galaxies. This likely due to the fact that edge-on galaxies are sampled by less pixels, which implies that it is harder to get robust shear measurements. Also, the \Ha\ emissions are extincted by dust more severely in edge-on galaxies and are less likely to be included~\citep{H09}. These selection also effects pose an interesting optimization avenue for a suitable KL sample. We leave a more rigorous discussion on sample selection effects to future work.

(2) For each $\bm{X}_\mrm{model}$ realization, we calculate a data vector of observables $\bm{Y}$=($\hat{\bm{\epsilon}}$, $v_\mrm{major}^\prime$, $v_\mrm{minor}^\prime$, $M_\mrm{B}$) out of equation~(\ref{eqn:estimators_image_to_source}), (\ref{eqn:galaxy_frame_source_plane_velocity}) and (\ref{eqn:TFR}). Here we assume a constant galaxy rotation curve $v_\mathrm{circ}=v_\mrm{TF}$. Then we apply measurement noise.
The noise is determined as follows:
    i) We assume shape measurement error $\sigma_{\bm{\epsilon}}$=0.009 per $\hat{\bm{\epsilon}}$ component, which is derived from equation~(13) in~\cite{Chang13} according to the typical photometric SNR and angular size of our KL sample.
    ii) The fiducial velocity measurement error is $\sigma_{v_\mrm{major}}$ = $\sigma_{v_\mrm{minor}}$ = 15 km$\,$s$^{-1}$. This velocity error is motivated by~\cite{OC20}, as stated in Sect.~\ref{sec:shape_noise}.
    iii) We assume a photometric SNR of 40 according to the typical KL sample selected from CMC (see Sect.~\ref{sec:sampleselection_KL}). The TFR intrinsic scatter is $\sigma_\mrm{TF}$=0.049 dex~\citep{R11}, and is absorbed into photometric measurement noise $\sigma_M=\sqrt{(1.0857/\mrm{SNR})^2+(\sigma_\mrm{TF}/b)^2}$.

(3) A multivariate Gaussian likelihood function is constructed for the model, along with a prior on $e_\mrm{int}$ as equation~(\ref{eqn:eint_prior}). We assume a diagonal covariance matrix for the data vector, with the variance determined in the previous step.

(4) For each realization, we optimize the log-likelihood function using the \textsf{scipy.optimize.minimize} routine starting around the neighbourhood of the fiducial model parameter. We also calculate the Fisher matrix via five-point stencil finite difference, then invert the matrix to estimate $\sigma^2(g_{1/2})$ at this parameter point.

(5) We evaluate the $\sigma^{-2}(g_{1/2})$-weighted ensemble mean and std, and take the weighted std as the shape noise, which is $\sigma(g)\approx0.025$. 

In the lower panel of Fig.~\ref{fig:rand_sanple}, we show the shape noise as a function of $e_\mrm{int}$. Source galaxies with smaller inclination angles have smaller shape noise and are more sensitive to the shear field than galaxies with larger inclination angles. Since our KL sample is biased towards low inclination galaxies, the weighting scheme is essential to suppress the effective shape noise per galaxy. 

%%%%%%%%%%%%%%%%%%%%%%%%%%%%%%%%%%%%%%%%%%%%%%%%%%

% Don't change these lines
\bsp	% typesetting comment
\label{lastpage}
\end{document}